**SUPERHUMPS IN CATACLYSMIC BINARIES.**

**XXV. $Q_{CRIT}$, ε(*Q*), AND MASS–RADIUS**

JOSEPH PATTERSON,[1] JONATHAN KEMP,[2,1,3] DAVID A. HARVEY,[4] ROBERT E. FRIED,[5]

ROBERT REA,[6] BERTO MONARD,[7] LEWIS M. COOK,[8] DAVID R. SKILLMAN,[9]

TONNY VANMUNSTER,[10] GREG BOLT,[11] EVE ARMSTRONG,[12,1] JENNIE MCCORMICK,[13]

THOMAS KRAJCI,[14] LASSE JENSEN,[15] JERRY GUNN,[16] NEIL BUTTERWORTH,[17]

JERRY FOOTE,[18] MARC BOS,[19] GIANLUCA MASI,[20] & PAUL WARHURST[21]

[1] Department of Astronomy, Columbia University, 550 West 120th Street, New York, NY 10027; jop@astro.columbia.edu
[2] Joint Astronomy Centre, University Park, 660 North A'ohōkū Place, Hilo, HI 96720; j.kemp@jach.hawaii.edu
[3] Visiting Astronomer, Cerro Tololo Interamerican Observatory, National Optical Astronomy Observatories, which is operated by the Association of Universities for Research in Astronomy, Inc. (AURA) under cooperative agreement with the National Science Foundation
[4] Center for Backyard Astrophysics (West), 1552 West Chapala Drive, Tucson, AZ 85704; dharvey@comsoft-telescope.com
[5] Center for Backyard Astrophysics (Flagstaff), Braeside Observatory; (deceased)
[6] Center for Backyard Astrophysics (Nelson), 8 Regent Lane, Richmond, Nelson, New Zealand; reamarsh@ihug.co.nz
[7] Center for Backyard Astrophysics (Pretoria), Post Office Box 11426, Tiegerpoort 0056, South Africa; lagmonar@csir.co.za
[8] Center for Backyard Astrophysics (Concord & Pahala), 1730 Helix Court, Concord, CA 94518; lew.cook@gmail.com
[9] Center for Backyard Astrophysics (East), 9517 Washington Avenue, Laurel, MD 20723; dskillman@comcast.net
[10] Center for Backyard Astrophysics (Belgium), Walhostraat 1A, B-3401 Landen, Belgium; tonny.vanmunster@cbabelgium.com
[11] Center for Backyard Astrophysics (Perth), 295 Camberwarra Drive, Craigie, Western Australia 6025, Australia; gbolt@iinet.net.au
[12] Department of Physics, University of California at San Diego, Mail Code 0354, 9500 Gilman Drive, San Diego, CA 92093; earmstrong@physics.ucsd.edu
[13] Center for Backyard Astrophysics (Pakuranga), Farm Cove Observatory, 2/24 Rapallo Place, Farm Cove, Pakuranga, Auckland, New Zealand; farmcoveobs@xtra.co.nz
[14] Center for Backyard Astrophysics (New Mexico & Uzbekistan), 9605 Goldenrod Circle, Albuquerque, NM 87116; loukrajci@comcast.net
[15] Center for Backyard Astrophysics (Denmark), Søndervej 38, DK-8350 Hundslund, Denmark; teist@image.dk
[16] Center for Backyard Astrophysics (Illinois), 1269 North Skyview Drive, East Peoria, IL 61611; jgunn@mtco.com
[17] Center for Backyard Astrophysics (Townsville), 24 Payne Street, Mount Louisa, Queensland 4814, Australia; neilbutt@bigpond.com.au
[18] Center for Backyard Astrophysics (Utah), 4175 East Red Cliffs Drive, Kanab, UT 84741; jfoote@scopecraft.com
[19] Center for Backyard Astrophysics (Otahuhu), Mount Molehill Observatory, 54 McDowell Cress, Glenfield, North Shore City, Auckland, New Zealand; molehill@ihug.co.nz
[20] Center for Backyard Astrophysics (Italy), Via Madonna de Loco, 47, 03023 Ceccano FR, Italy; gianluca@bellatrixobservatory.org
[21] Department of Physics, University of Auckland, Private Bag 92019, Auckland, New Zealand; p.warhurst@auckland.ac.nz

## ABSTRACT

We report on successes and failures in searching for positive superhumps in cataclysmic variables, and show the superhumping fraction as a function of orbital period. Basically, all short-period systems do, all long-period systems don't, and a 50% success rate is found at $P_{\rm orb}$=3.1±0.2 hr. We can use this to measure the critical mass ratio for the creation of superhumps. With a mass–radius relation appropriate for cataclysmic variables, and an assumed mean white-dwarf mass of 0.75 $M_\odot$, we find a mass ratio $q_{\rm crit}$=0.35±0.02.

We also report superhump studies of several stars of independently known mass ratio: OU Virginis, XZ Eridani, UU Aquarii, and KV UMa (= XTE J1118+480). The latter two are of special interest, because they represent the most extreme mass ratios for which accurate superhump measurements have been made. We use these to improve the ε($q$) calibration, by which we can infer the elusive $q$ from the easy-to-measure ε (the fractional period excess of $P_{\rm superhump}$ over $P_{\rm orb}$). This relation allows mass and radius estimates for the secondary star in any CV showing superhumps. The consequent mass–radius law shows an apparent discontinuity in radius near 0.2 $M_\odot$, as predicted by the disrupted magnetic braking model for the 2.1–2.7 hour period gap. This is effectively the "empirical main sequence" for CV secondaries.

## 1. INTRODUCTION

Superhumps are large-amplitude periodic waves in the light curves of dwarf novae in superoutburst, and several other types of cataclysmic and low-mass X-ray binaries. Their most curious hallmark is their *period* — which is typically a few percent longer than the orbital period, but changes slightly during the course of an outburst. There is now a complex zoology of related periodic phenomena (Patterson et al. 2002a, Appendix A), although most stars participating in this phenomenon are dwarf novae and show "common" superhumps which evolve in a consistent and repeatable way — substantially as described in the earliest papers on the subject (Vogt 1974; Warner 1975, 1985).

A series of theory papers in 1988–1992 established the reason for superhumps: an eccentric instability grows at the 3:1 resonance in the accretion disk, and perturbation by the secondary forces the eccentric disk to precess (Whitehurst 1988, Hirose & Osaki 1990, Lubow 1991). Recent studies have confirmed this, added new details, and established the basic origin of the superhump light: the extra heating associated with periodic deformations of disk shape (Lubow 1992; Murray 1996, 2000; Kunze et al. 1997; Simpson & Wood 1998; Wood, Montgomery, & Simpson 2000). These studies have found that the mass ratio $q$=$M_2$/$M_1$ plays a key role, with superhumps produced only below a critical ratio, roughly $q_{\rm crit}$~0.3. This limit has become an article of faith; but different models give different estimates, and it has never been established by observations of disks in actual stars (rather than in computers).

The perturbation making superhumps should scale with $q$, and it has therefore been tempting to use the observed precession rates as estimators of $q$, $M_2$, or $M_1$. This is a worthy goal, since the latter quantities are of great interest and are famously difficult to measure. The earliest such efforts, starting with Mineshige et al. (1992), achieved only limited success, because they relied on expressions for $q_{\rm crit}$ and ε($q$) derived solely from theory. Even to this day, there is still no published work which provides an evidentiary basis for knowing the value of $q_{\rm crit}$. A recent study (Patterson 2001, hereafter P01) proposed to replace the theoretical ε($q$) dependence [roughly ε=0.34$q$] with an empirical expression [ε=0.22$q$] which correctly reproduces the (ε, $q$) values for the best-studied stars. But this expression was only well-calibrated for middling values of $q$ (~0.10–0.17). In this paper, we report new observations at low and high $q$ to remedy this shortcoming, and study the dependence of superhumps on $P_{\rm orb}$, to establish a measured $q_{\rm crit}$. We also show how ε($q$) can be used to establish an accurate mass–radius relation for the secondaries in short-period cataclysmic variables.

## 2. OBSERVATIONS

This paper is another in our series on the superhumps of cataclysmic binaries (just in case you didn't know). As in previous papers, we study long time series of differential photometry in order to identify periodic signals. Periods displaced by a few percent from $P_{\rm orb}$ are identified as superhump phenomena. Most of the light curves were obtained at observing stations of the Center for Backyard Astrophysics (CBA), a worldwide network of small telescopes (http://cba.phys.columbia.edu/). A wide distribution in longitude is especially important, because period-finding can be frequently confused by 24-hour aliasing.

In this paper we report details of photometric campaigns on five stars, and discuss three others — all stars with independently known mass ratios. This is in order to establish an empirical $\varepsilon(q)$ calibration. Then we will report our overall survey results and try to answer the question "which stars superhump?" We will find that superhumping is strongly correlated with orbital period — sufficiently strongly that the mass ratio may be the sole determining factor. These are essentially *summary* reports on the individual observing campaigns; some will be reported later in greater detail.

## 3. OU VIRGINIS

The history of this eclipsing dwarf nova is well reviewed by Vanmunster, Velthuis, & McCormick (2000). Superoutbursts occurred in May 1999, June 2000, May 2002 (probably), and May 2003. Thus the recurrence interval could be ~360 days. However, the superoutbursts of *many* dwarf novae recur with similar periods, so this probably reflects the inevitable periodicity of visual monitoring activity for stars near the ecliptic. The star is still not extensively monitored, and superoutbursts last only 10–14 days, so some superoutbursts (and *most* short outbursts) will likely be missed.

A superoutburst was discovered and announced by R. Stubbings and M. Simonsen on 4 May 2003 (JD 2452763). CBA photometry began 2 nights later, with the star near $V$=15.1, and continued for 50 nights. The upper left frame of Figure 1 shows the eruption light curve, and a summary observing log is given in Table 1. The eruption light curve looks altogether normal for SU UMa-type dwarf novae: an unobserved rise, a steady decline at ~0.12 mag/day, followed by a sharp drop to a level slightly above minimum light. That level was near $V$=17.6, about 0.5–1.0 magnitude above true quiescence; this too is a common feature of these dwarf novae.

All the light curves were punctuated by strong superhumps, and deep eclipses recurring with $P_{\text{orb}}$=0.07271 d. The eclipse depths varied from 0.4 mag to 1.0 mag on the "precession" cycle (the beat of orbital and superhump periods), as is also commonly found in eclipsing dwarf novae. The mean orbital waveform near maximum light (summed over the first five days, to encompass exactly two beat cycles) is shown in the upper right frame of Figure 1.

We obtained very extensive coverage for the first 8 days of outburst. To prepare the light curves for time-series analysis, we subtracted the mean and trend from each night's light curve, spliced them together, and removed eclipses. The resultant light curve, seen in the middle frame of Figure 1, is dense enough to overcome aliasing difficulties, and long enough to give good frequency resolution. The bottom frame of Figure 1 contains the resultant power spectrum, showing an obvious strong signal at the superhump frequency $\omega_{\text{sh}}$=13.32(2) c/d and two harmonics.[22] Synchronous summation at that frequency gave the mean waveform, shown in the inset. This is basically the familiar waveform of a *common superhump*.

We then subtracted these signals from the time series and recalculated the power spectrum, in order to look for weaker periodic signals which might be hiding in the window pattern of the superhump. We found no certain detections, although there were peaks at 40.84 and 53.68 c/d (not shown), which correspond respectively to $3\omega_o-\Omega$ and $4\omega_o-3\Omega$ under the usual $\omega_{sh}=\omega_o-\Omega$ prescription (where Ω is the putative — and generally unseen — frequency of disk precession).

[22] In this paper we use c/d as a shorthand for cycles day$^{-1}$.

The superhump seemed to endure throughout the 50-day campaign, although we could not unambiguously track it past the first 16 days. Just for the record, its ephemeris near eruption peak was found to be

$$\text{Maximum light} = \text{HJD } 2{,}451{,}765.1201(10) + 0.07508(9)\, E. \qquad (1)$$

The observed fractional period excess of the superhump is $\varepsilon=(P_{sh}-P_{orb})/P_{orb}=0.0326(15)$. Resolution of the eclipse shape at quiescence has yielded a fairly accurate value for the mass ratio (Feline et al. 2004a). This will be useful in Section 12.

## 4. XZ ERIDANI

Woudt & Warner (2001) reported deep eclipses in the quiescent light curves of this dwarf nova, recurring with a period of 0.06116 d. This made the star an attractive target for photometry during outburst. So when Rod Stubbings found the star at *V*=15 on 27 January 2003, we immediately launched a campaign, which continued for 10 days. The first observations, those of Greg Bolt on 28 January, showed obvious strong superhumps. Much of our data has already been included in the study of Uemura et al. (2004); here we give a briefer description limited to the superhumps. The summary observing log is given in Table 2.

A sample light curve is given at the top of Figure 2, which shows large superhumps and sharp eclipses. We obtained good global coverage (Africa, North America, New Zealand, Australia), and hence were essentially immune to aliasing problems. The power spectrum of the spliced 10-day light curve, with eclipses removed, is given in the lower frame of Figure 2, with significant signals flagged with their frequencies. The main signal occurs at $\omega_{sh}$=15.915(11) c/d, along with the first two harmonics. In addition there are significant signals slightly blue-shifted from the harmonics. In the customary $\omega_{sh}=\omega_o-\Omega$ notation, these signals occur at $2\omega_o-\Omega$ and $3\omega_o-2\Omega$. These "sideband" signals are commonly produced in CV superhumps. Inset is the mean superhump waveform, which tracked the ephemeris

$$\text{Maximum light} = \text{HJD } 2452668.0336 + 0.06283(8)\text{E}. \qquad (2)$$

The superhump period excess is ε=0.0273(13). Feline et al. (2004b) analyzed white-dwarf and hot-spot contact times in the quiescent eclipses, and derived an accurate *q*=0.1098(17). These numbers will be useful in Section 12.

## 5. UU AQUARII

This novalike variable eclipses with a period of 0.16358 d, and has been well studied by Baptista et al. (1994), who deduced *q*=0.30(7) from the eclipses. This makes it of high interest for superhumps. A 1998 CBA observing campaign gave some evidence for periods slightly longer than $P_{orb}$; but the signals were weak, somewhat unstable, and troubled by aliasing. So we improved our density and distribution of coverage, and tried again in 2000. This campaign covered 57 "nights" and 270 hr of data, spread over 21 days. The summary log is given in Table 3. Observing stations spanned the full range of terrestrial longitude (North America, New Zealand, Australia, Europe), so there was no confusion due to aliasing. And the star cooperated by staying in a single luminosity state ($V$~13.9) throughout the campaign.

Figure 3 shows a power spectrum of the full light curve, after removing eclipses. A powerful superhump is present at 5.711(6) c/d, along with two harmonics. In addition there is a satellite signal at 17.565(10) c/d. Adopting $\omega_o$=6.113 c/d and the usual convention that $\omega_{sh}=\omega_o-\Omega$, we see that the satellite signal occurs at $3\omega_o-2\Omega$. This fine-structure is a common property of apsidal superhumps (Skillman et al. 1999). Just for the record, the superhump tracked the ephemeris

$$\text{Maximum light} = \text{HJD } 2451790.1013 + 0.17510(18)\, E. \qquad (3)$$

## 6. KV URSAE MAJORIS (= XTE J1118+480)

This star, now an extensively studied low-mass X-ray-binary, was discovered as a bright transient by the Rossi X-ray Timing Explorer on 29 March 2000 (Remillard et al. 2000). CBA photometry began two days later, and the first night's observation (by Lew Cook) showed a 4-hour signal in the light curve, which proved to be stable after a few days' coverage (Cook et al. 2000, Patterson et al. 2000a). We continued the campaign for 88 days, accumulating a total of 460 hrs over 92 nights. The summary observing log is given in Table 4.

Figure 4 shows the power spectrum of the first 30 days, revealing a stable signal at 5.857(5) c/d. The mean waveform, shown in the inset figure, is essentially a pure sinusoid. Maximum light tracked the ephemeris

$$\text{Maximum light} = \text{HJD } 2451634.848 + 0.17073(15)\, E. \qquad (4)$$

We tracked this signal throughout the campaign. The amplitude and waveform remained substantially unchanged, but timings of maximum and minimum light gradually drifted from a constant-period ephemeris. Figure 5 shows an O–C diagram of the timings during the first 45 days of our campaign, relative to a test period of 0.17065 d. The curvature indicates a slow period decrease ($\dot{P}$ =–1×$10^{-6}$). Table 4 catalogues the period changes in various intervals.

More details about the superhump are given by Uemura et al. (2002) and Zurita et al. (2002). The orbital period is known to be 0.169937(1) d from the observed ellipsoidal modulation of the secondary in quiescence (Zurita et al. 2002), and we adopt $P_{sh}$=0.17073(12) d as a mean value appropriate to the eruption peak (within 0.4 mag of maximum light, the same convention we use in judging $P_{sh}$ in dwarf novae). Thus we obtain ε=0.0047(9). The mass ratio is more elusive, since the system does not eclipse; but Orosz et al. (2001) argued for $q$=0.037(7) from measuring the rotational velocity of the secondary.

This supplies another point for the ε($q$) calibration in Section 11. The inclusion of a black-hole X-ray transient in this collection of CVs may be problematic. But superhumps have consistently been seen in the optical light curves of X-ray transients, and their properties bear a reasonable resemblance to those of CVs (O'Donoghue & Charles 1996, Zurita et al. 2002, Figure 1 of P01). So for the present, we accept this calibration point.

From the earliest observation, we were puzzled that the light curves were so noisy, despite no obvious flickering. Essentially all light curves showed a 0.1 mag noise band. After a few days of bafflement, we tried fast photometry at 2 s time resolution; this still showed a large noise band, but power spectra of short segments revealed significant peaks at periods near 10 s. The peaks moved around rapidly, and we only achieved a stable result by averaging power spectra of many short observations. The left frame of Figure 6 shows the mean power spectrum from 36 10-minute observations on JD 2451665. Other detections of this quasi-periodic signal are listed in Table 4, and shown in the right frame of Figure 6.

## 7. BB DORADUS (= EC 05287–5857)

This novalike variable was discovered in the Edinburgh–Cape survey for blue stars (Chen et al. 2001). Chen et al. suggested a tentative period of 0.107(7) d in a radial-velocity search, and this motivated us to carry out a photometry campaign in October–November 2002. We accumulated 140 hr of coverage over 28 nights during the 45-day campaign. The star remained near $V$=14.0 throughout, the "high state" of this probable VY Scl star. A summary observing log is given in Table 5.

The light curves were typical of most novalikes, dominated by strong flickering; samples are given in Figure 8 of Chen et al. The top frame of Figure 7 shows the average ("incoherent") nightly power spectrum, formed by averaging the nights of long (>6 hr) coverage. A signal near 6.7±1.0 c/d is present, and flickering with $P\propto\nu^{-1.3}$. The middle frame shows the "cleaned" power spectrum (alias peaks removed by subtraction) of the 45-day light curve, with the frequencies of significant signals marked. There was no detection near the candidate orbital frequency. The longitude distribution was essentially perfect (Chile, New Zealand, Australia, South Africa), so there were no aliasing problems. This enabled us to clean the power spectrum without ambiguity.

The dominant signal occurs at 6.701(3) c/d, with weaker signals at 6.126 and 12.833 c/d. This trio of frequencies suggests a familiar pattern in the numerology of superhumps: if the strongest signal is $\omega_o$, then the other signals could be superhumps at $\omega_o-\Omega$ and $2\omega_o-\Omega$. The mean orbital and ($\omega_o-\Omega$) superhump waveforms are seen in the lowest frame of Figure 7. Since this pattern is very common and the radial-velocity evidence is quite weak, we are willing to bet on this identification. The orbital ephemeris is then

$$\text{Maximum light} = \text{HJD } 2452583.438(6) + 0.14923(7)\,E. \qquad (5)$$

But we might be wrong. It is also possible that the dominant signal is a 'negative' superhump, with the wave blue-shifted to a higher frequency $\omega_o$+N, and the weaker signal at 6.126 c/d representing $\omega_o$. If so, then the third signal occurs at $2\omega_o+N$ — occasionally, though not commonly, seen among negative superhumpers. The main reason we disfavor this interpretation is that the required value of ε (–0.094) is somewhat out of bounds for the family of negative superhumpers (see Figure 1 of Patterson 1999), The equivalent value for a *positive* superhump interpretation [ε=0.094(1)] accords fairly well with what is expected at such a long $P_{\rm orb}$ (see Sec. 10–12 below).

Naturally, a radial-velocity study to resolve this ambiguity is warmly recommended!

## 8. U GEMINORUM AND IP PEGASI

These two dwarf novae are of special interest because they are eclipsing double-lined binaries, thereby satisfying in principle the requirement for accurate measurement of masses. Smak (2001, 2002) deduced $q$=0.36(2) and 0.45(4) for these stars respectively. Extensive CBA photometry has failed to reveal superhumps in either star — even in very long eruptions, as long as the superoutbursts of SU UMa stars.[23] The U Gem upper limit on $q$ may be particularly constraining, because several superhumping stars have a $q$ provably or likely near 0.3. Below we will use this, in conjunction with other data, to estimate an upper limit of $q_{\rm crit}$=0.35 for the creation of superhumps.

## 9. DW URSAE MAJORIS

DW Ursae Majoris is an eclipsing novalike variable which shows high and low states, and many phenomena charactistic of the "SW Sex" class of CVs. A fortuitous HST observation during a low state revealed the steep-walled eclipse of a pure white dwarf, enabling the deduction of a $q(i)$ relation (Araujo-Betancor et al. 2003). This yielded $q$>0.24. We have reported CBA time-series photometry, which shows long-lived episodes of negative and positive superhumps, yielding ε=0.0644(20) (Patterson et al. 2002b). Since we know of no star showing superhumps at $q$>0.35, and U Gem apparently fails to superhump at $q$=0.36±0.02, we are inclined to regard 0.36 as an upper limit. Thus we regard $q$=0.24–0.36 as being very likely in DW UMa — sufficiently likely to use as a constraint.

Also, under the assumption that ε is strictly a function of $q$, DW UMa is probably not near the upper limit of $q$ — since we need to leave room for stars with larger ε. Let us adopt $q$=0.24–0.32. For this $q$, Araujo-Betancor et al. (2003) deduced a white-dwarf mass of 0.73(3) $M_\odot$ to fit the observed durations of ingress and egress. This determines $M_2$ and $R_2$ from Kepler's Laws: from (0.18 $M_\odot$, 0.29 $R_\odot$) to (0.24 $M_\odot$, 0.32 $R_\odot$). These are not exactly "hard" measurements, since they require an assumption from superhump theory. But it's a fairly weak assumption (that ε depends solely or mainly on $q$); and in an enterprise where constraints are hard to find, we consider them worthy of use.

23 A recent study of AAVSO visual data on the long U Gem outburst of 1985 appears to contradict this (Smak & Waagen 2004). Our data so far, including some in the same outburst, does not confirm this; and the published evidence does not seem strong. But the point definitely warrants closer study.

## 10. WHICH STARS SUPERHUMP?

We have been measuring CV light curves for superhumps since 1977, and systematically since 1993. To date we have obtained results for ~200 stars. At first we concentrated on short-period dwarf novae, since the first superhump discoveries were in such stars. Soon we expanded the search to include all CVs bright enough (usually *V*~15–16) to get onto our radar screens. In a typical case, we use the globally distributed telescopes of the CBA for a campaign lasting 1–2 weeks, with ~40–150 hr of observation. Power-spectrum analysis of the light curve then tests for the presence of periodic signals. The detection threshold is always sensitive to the erratic variability ("flickering") for which CVs are famous. Dwarf novae in outburst typically have very quiet light curves, in which case 1–3 hour periodic signals can be detected down to a semi-amplitude of ~0.015 mag. Novalike variables typically have strong flickering, which degrades the detection limit to ~0.03 mag.

Other groups (especially the Kyoto group: Kato et al. 2004) have also been publishing superhump detections, of course. But so far there has been no accounting of the actual success rate, or of the observing strategy. Thus the perceptions of superhumps in the greater CV and accretion-physics communities have been in the form of "conventional wisdom", gradually formed by sheer weight of publication rather than addressed in any specific study. We aim to remedy that situation here.

As stated above, we tend to target all CVs above our brightness threshold. But very early in this enterprise, it was clear that superhumps — or at least the *common superhumps* which are the subject of this paper — are absent in quiescent dwarf novae. It was also apparent that superhumps avoid CVs of long orbital period, and avoid highly magnetic CVs (AM Her stars). Of course, these are three elements of today's conventional wisdom. Since we wished to test the latter, we invested considerable time in stars which would violate those rules. But since we also wished to discover superhumps, we used those rules somewhat in target selection, especially when the rules seemed to be very firm indeed. So our study does not quite have the merit of an "unbiased survey". As a result of this dual strategy, we found many common superhumps, but also accumulated enough information about the range of the phenomenon to write this section of our paper.[24]

Most of the brighter CVs have a known orbital period. Since superhumps obviously favor short $P_{\rm orb}$, we tried to observe all CVs with $P_{\rm orb}$<4 hr. For old novae and novalikes, we succeeded (within our brightness limits). Some dwarf novae eluded our net by virtue of rare outbursts or southerly declination; but we still managed to observe ~120 dwarf novae with $P_{\rm orb}$<4 hr. We also observed a few dozen stars of longer $P_{\rm orb}$, and a few dozen of unknown $P_{\rm orb}$.

We have published superhump detections in ~60 stars, with the details of ~50 more detections still awaiting a precise measure of $P_{\rm orb}$. Table 6 summarizes the overall results, with most of the positive detections previously tabulated (Patterson 1998, Patterson et al. 2003;

[24] And to find a range of phenomena related to superhumps but much more poorly known. A one-page summary of these can be found in Appendix A of Patterson et al. (2002a).

hereafter P98, P03). Nearly all short-$P_{orb}$ stars in their "high states" show superhumps, and all long-$P_{orb}$ stars don't. For dwarf novae, the brightness is the obvious high-state identifier. Most old novae and novalikes are considered to be in a high state, because their absolute magnitudes are similar to those of erupting dwarf novae ($M_V$~4–6); a few are much fainter, and none of the latter show superhumps. Thus a better statement is probably: *all nonmagnetic CVs superhump if they have a $P_{orb}$ sufficiently short and an $\dot{M}$ sufficiently high*. How short, and how high?

Based on Table 6, Figure 8 shows the dependence of superhump percentage on $P_{orb}$. That percentage declines to 50% at $P_{orb}$=3.1±0.2 hr, and no (positive) superhumps are definitely seen with $P_{orb}$>4 hr. A few other notes are pertinent to this result.

1. Table 6 — and for that matter all remaining discussion — applies to all (nonmagnetic) CVs, not just those with CBA data. The difference is minor, though; we obtained adequately long time series for 126 of the 138 stars with positive detections, and all 68 nondetections. We did exclude from consideration ~6 stars for which superhump claims have been made, when we judged the quality of evidence to be insufficient (usually because adequate power-spectrum evidence is not presented).

2. We mainly excluded AM Her stars from consideration, but observed all the sufficiently bright DQ Her stars (intermediate polars). None of the latter produced clear superhumps, although the reason for this is unclear ("high state" is poorly defined for such stars).

3. A possible exception to these trends is TV Col, a magnetic CV where Retter et al. (2003) reported positive superhumps despite the long $P_{orb}$ (5.5 hrs). Our two campaigns on TV Col did not confirm this; but superhumps are known to be transient in many novalikes, so the star remains an interesting candidate. Unfortunately, it is unlikely that magnetic CVs can ever be used for our purposes [$q_{crit}$, ε($q$), mass–radius], since they do not contain full disks.

4. Although superhumps are clearly dependent on a high–$\dot{M}$ state, the causal relation is unclear. It seems likely that the true underlying cause of both is viscosity, which permits accretion and enlarges the disk, thereby giving access to the resonance. Since $\dot{M}$ is (roughly) measurable and viscosity is not, we characterize the dependence in terms of $\dot{M}$. That might be correct. But if — as is likely — the real requirement is the transition to the thermal instability, then the relevant condition is more complex. Osaki (1996) characterized the thermal instability condition as roughly $\dot{M}_{crit}=10^{-9}\ M_{\odot}/\mathrm{yr}\ P_o\ [\mathrm{hr}]^{1.8}$, and that seems about right.

5. These are the circumstances relevant to superhump *manufacture*. But superhumps commonly outlive the high–$\dot{M}$ states which spawn them, sometimes for as long as a few hundred or a few thousand orbits. No one quite understands this.

6. Are there superhumps just below our detection threshold? Well, we *never* found common superhumps of dwarf novae near the threshold; all were much stronger, or were undetected. Thus we suspect that our net was 100% accurate for dwarf novae — or to put it another way, "there are no small common superhumps". For novalikes, too, most detections were far above threshold; but a few were not, and a few others were classified as nondetections because they barely failed to have sufficient amplitude and/or repeatability. We are confident that all detections are certain, but a few classified nondetections may simply result from low sensitivity.

## 11. THE TRANSITION REGION NEAR 3 HOURS

In a perfect world, superhumps would depend strictly on $q$, all white dwarfs would have the same mass, and all secondaries would obey a single mass–radius relation (e.g., "the main sequence"). If all of that were true, there would be a perfectly sharp transition separating the superhump-eligibles from the ineligibles. There would also be a perfect correlation between $P_{\text{orb}}$ and ε.

Just how perfect is our world? Well, in this very limited sense, it seems to be pretty good. The rms scatter in $\varepsilon(P_{\text{orb}})$ is quite low, just 21% (see Figure 20 and Sec. 6 of P03). This implies a scatter of <21% in $M_1$ and <11% in mass–radius. We can also apply superhump theory to the $P_{\text{orb}}$ dependence seen in Figure 8 to obtain a constraint on $q_{\text{crit}}$. Applying $P_{\text{orb}}$=3.1±0.2 hrs to Eqs. (7), (8), and (9) of P03, we obtain

$$q_{\text{crit}} = (0.36 \pm 0.03) \langle m_1 \rangle^{-1} \alpha^{-2.05} \qquad (6)$$

from Figure 8, where $\langle m_1 \rangle = \langle M_1 \rangle / M_\odot$ and $\alpha = R_2/R_{\text{ZAMS}}$. Since the observed upper limit to $q$ is very nearly the same (0.32–0.38, see below), we basically have

$$\alpha = \langle m_1 \rangle^{-1/2}. \qquad (7)$$

This is nearly the same constraint as obtained in P03 from a separate argument (the value of ε at a given $q$, see Sec. 6.2 and 6.3 of P03). For an assumed $\langle m_1 \rangle$=0.75, it implies α=1.15 — secondaries 15% larger than $R_{\text{ZAMS}}$.

What about the *width* of the transition zone near 3.1 hr? We don't have many stars in this region, but the narrowness of the transition implies that no great variance can exist in $M_1$ or α. We carried out numerical experiments on trial populations of CVs with widely varying values of $M_1$, and found that the sharpness of the transition could not be reproduced with an rms scatter in $M_1$ greater than 30%. This is consistent with, but not quite as constraining as, the result from the scatter in $\varepsilon(P_{\text{orb}})$ (discussed in P03).

## 12. ε(*Q*)

What value of ε is produced by a given $q$?

We previously tackled this question with the data available in 2000 (P01). That study contains a fuller discussion of these matters. In brief, we derived an empirical $\varepsilon(q)$ calibration from observations of eclipsing binaries of known $q$. A few new calibrating stars are now known, a few estimates are improved, and we now have an estimate for how the superhump phenomenon is distributed with $P_{\rm orb}$. These make it desirable to revisit the issue.

The relevant data are shown in Table 7 and Figure 9. The P01 linear $\varepsilon(q)$ scaling remains a *roughly* acceptable fit to the data. But ε ranges as high as 0.094 (at least), which implies $q$=0.43 on the P01 scaling. This is probably too high. Theoretical models have suggested $q_{\rm crit}$ in the range 0.20–0.35, and we assume the upper limit for U Gem (0.38) to be reliable. Refitting the data after assigning $q$<0.38 to the largest known ε (BB Dor), we find a scaling slightly different at large $q$:

$$\varepsilon = 0.18\, q + 0.29\, q^2. \tag{8}$$

This is likely an improvement on the P01 relation. On the assumption that ε is strictly a function of $q$, this allows use of the easily measured ε as a surrogate for the elusive $q$.

## 13. MASS–RADIUS RELATION FOR THE SECONDARIES

So superhumps, when observed with sufficient precision and accompanied by an accurate $P_{\rm orb}$, seem to yield an accurate $q$. The observed low dispersion in $\varepsilon(P_{\rm orb})$ (Figure 20 and Sec. 6 of P03) and to a lesser extent the sharpness of the transition in Figure 8 of this paper also demonstrate that the dispersion in $M_1$ does not exceed ~20%. This implies an estimate of $M_2$, not merely $q$, for each superhumping star. And since lobe-filling secondaries obey a $P\sqrt{\rho}$ relation, it also implies a radius for each star.

Let's see how this goes. The secondaries fill their Roche lobes, requiring

$$P_{\rm orb}\,[\rm hr] = 8.75\,(m_2\,/\,r_2^{3})^{-1/2}, \tag{9}$$

where $m_2$ and $r_2$ are in solar units (Faulkner, Flannery, & Warner 1972). This implies

$$r_2 = 0.2355\, P_{\rm orb}^{2/3}\, m_1^{1/3}\, q^{1/3}. \tag{10}$$

Figure 9 suggests that $q$ is probably deducible from ε within ~10% in its best calibrated region ($q$~0.10–0.20). This error might rise as high as 20% for the smallest and largest values of $q$. The resultant error in $r_2$ is small, just 3–7%. The error resulting from the dispersion in $m_1$ is also small (<7%, since $m_1$ varies by <20%). Unfortunately, though, superhump theory does not specify the *actual value* of $m_1$; it is coupled with α via Eq. (7).

Help in this matter comes from several sources. Conventional measures of $\langle m_1\rangle$ in CVs have yielded fairly consistent results: 0.7±0.1 (Shafter 1983), 0.74±0.04 (Webbink 1990), and 0.77±0.21 (Smith & Dhillon 1998). We consider $\langle m_1\rangle$=0.75 to be a good choice. Secondly, the eight accurately known mass–radius pairs in short-period eclipsing CVs show an α substantially exceeding 1 [~1.2, comparing Table 8 and Figure 10 with the theoretical ZAMS of Baraffe et al. (1998)]. And thirdly, CV evolution models (e.g., Kolb & Baraffe 1999) adopting α=1 always reach too short a minimum $P_{\rm orb}$ (~66 compared to the observed 78 minutes); since $r_2 \propto P_{\rm orb}^{2/3}$, this can be roughly fixed by adopting α=1.15. Thus the evidence favors $\langle m_1\rangle$=0.75, α=1.15,

implying

$$r_2 = 0.214\, P_{\rm orb}^{2/3}\, q^{1/3} \qquad (11)$$

and

$$m_2 = 0.75\, q. \qquad (12)$$

For those few stars with a better $m_1$ constraint available (eclipsers, and fast classical novae — for which we assign $m_1$=1), we used that constraint rather than the default $m_1$=0.75.

We then calculated the ($m_2$, $r_2$) pairs from the ε($q$) relation, and show them as the filled squares in Figure 10, along with values directly and independently obtained in eclipsing binaries (indicated by crosses). The data are shown in Table 9.[25] The solid curve is the theoretical ZAMS mass–radius relation (the 10 Gyr models of Baraffe et al. 1998), extended to lower mass with the cold brown dwarf models of Burrows et al. (1993). This is very similar to the correlation shown in Figure 2 of P01, but differs slightly for high $m_2$ (>0.18) because of the changed ε($q$) calibration.

The main lesson is the same one drawn by P01: CV secondaries are larger than ZAMS stars (i.e., α>1). Although the typical error in radius is ~10–12%, similar to the disagreement with the ZAMS, it is nevertheless certain that the CV secondaries lie above the ZAMS, because the error diamonds move the points roughly *parallel* to the ZAMS — and also because the independently derived crosses are well above the ZAMS.

An apparent discontinuity appears around 0.20 $M_\odot$. Secondaries less massive show great consistency, just slightly above the ZAMS (α~1.10). Secondaries more massive appear to be generally larger, with α~1.30. Stars with $P_{\rm orb}$>2.5 hr are identified with enclosing boxes. Thus these two categories define the two sides of the "period gap" — the $P_{\rm orb}$=2.1–2.7 hour region where very few stars are found.[26]

In the simplest understanding of the period gap, secondaries detach from their Roche lobes when they become completely convective (Robinson et al. 1981, Rappaport et al. 1983). For single ZAMS stars this occurs near 0.35 $M_\odot$ — wildly inconsistent with our estimate of 0.196±0.014 $M_\odot$ from Figure 10! However, stars losing mass are much bigger; for appropriate values of $\dot{M}$ (>$10^{-9}$ $M_\odot$/yr), the corresponding mass has been calculated as 0.255±0.015 $M_\odot$ (McDermott & Taam 1989, especially their Figure 1 and Table 1) and 0.23±0.01 $M_\odot$ (Howell et al. 2001, especially their Figure 2). Our measured value is slightly discrepant. But while the sharpness of the discontinuity in Figure 10 is significant (and constitutes some evidence for low dispersion in $m_1$), its location at 0.20 $M_\odot$ depends on the assumption that $<m_1>=0.75$. Since $m_2 \propto m_1$, a change to $<m_1>=0.9$ would move the discontinuity to 0.24 $M_\odot$. So within the errors of the observational data, and perhaps also the theory, there is no certifiable disagreement between these estimates.

[25] Most values of $q$ are given in Table A1 of Patterson, Thorstensen, and Kemp 2005; others can be deduced from the ε tabulations of P03 and P98. These papers give more complete references.

[26] For CV fans *in extremis*, the unboxed points of largest $m_2$ and $P_{\rm orb}$ are NY Ser and V348 Pup — stars with $P_{\rm orb}$=2.35 and 2.44 hr respectively. There is some sensitivity here — certainly in our discussion, and probably in the actual binaries as well — to exactly where the edges of the period gap are placed.

The most popular theory for the origin of the period gap — disrupted magnetic braking (Ritter 1985; Taam & Spruit 1989; Hameury, King, & Lasota 1991; Spruit & Ritter 1983) — predicts such a discontinuity. The radii of stars immediately above and below the period gap ($R_a$ and $R_b$) should be in the ratio

$$R_a / R_b = (P_a / P_b)^{2/3}, \tag{13}$$

where $P_a$ and $P_b$ are the periods above and below the gap [from Eq. (9) above]. For a 2.1–2.7 hour gap, we thus expect an 18% difference — in agreement with the effect estimated from Figure 10 (1.30/1.10 implying a 18% difference). Figure 10 thus offers substantial evidence in support of the disrupted magnetic braking model, or models like it which predict bloated secondaries above the period gap.

Figure 11 presents another version of this data, with $R_2/R_{\rm ZAMS}$ averaged over 0.01 and 0.02 $M_\odot$ bins. This does not add anything new, but perhaps better illustrates the trends discussed here: the discontinuity at 0.20 $M_\odot$, and the progressive bloating of the secondary as it evolves to very low mass.

Figure 12 shows the mass–radius data on a log–log scale, including all additional eclipsing CVs with $P_{\rm orb}$<8 hr (also contained in Table 8).[27] Just for the record, the resultant empirical mass–radius relation for CV secondaries is

$$r_2 = \begin{cases} 0.62\, m_2^{0.61} & [m_2 = 0.06\text{–}0.18,\ P_o=1.3\text{–}2.3] \\ 0.92\, m_2^{0.71} & [m_2 = 0.21\text{–}1.0\ ,\ P_o=2.5\text{–}8.0] \end{cases} \tag{14}$$

This relation is shown as the piecewise linear fit in Figure 12. The equivalent mass–period relation is

$$m_2 = \begin{cases} 0.032\, P_o^{2.38} & [P_o < 2.3] \\ 0.026\, P_o^{1.78} & [P_o > 2.5] \end{cases} \tag{15}$$

The data are too sparse to yield trustworthy results for the secondaries of lowest mass $m_2$<0.06).

[27] Restriction to eclipsers improves reliability over some previous collections, e.g., Table 3 of Smith & Dhillon (1998). Noneclipsing CVs seldom offer an inclination constraint good enough to use in these calibrations — except in superhump studies, where inclination is irrelevant.

These relations could be considered the "empirical main sequence" of CVs. The low dispersion in Figures 10 and 12, like that of the more famous main sequence, requires a fairly uniform chemical composition in these secondaries. In particular, stellar radius is sensitive to mean molecular weight, so these secondaries probably have not burned appreciable hydrogen in their prior lives — either before or after their imprisonment in a cataclysmic binary. To a good approximation, they should be considered unevolved stars of solar composition.[28]

There are a few obvious exceptions to this. A few secondaries (AE Aqr and GK Per are famous examples) have radii much too large for their mass, apparently as a result of true nuclear evolution. All of these seem to be in binaries with $P_{\rm orb}$>8 hr (hence outside our net). And three others have radii much too small for their mass; these are all known or suspected to be helium-rich (EI Psc: Thorstensen et al. 2002a; QZ Ser: Thorstensen et al. 2002b; V485 Cen: Augusteijn et al. 1996). But the great majority of CV secondaries seem to be pretty normal low-mass stars — just slightly agitated by the special circumstances of their imprisonment (enforced rapid rotation, struggles with thermal equilibrium, the occasional nova outburst, etc.).

## 14. SUMMARY

1. We report photometric campaigns on the eclipsing dwarf novae XZ Eridani and OU Virginis. These showed common superhumps with ε=0.0270(15) and 0.0326(15), respectively, for the two stars.

2. We do the same for the novalike variable UU Aquarii. The 2000 observing campaign showed strong superhumps — a stable wave with ε=0.0702(24), lasting essentially throughout the 50-day campaign.

3. We discovered and tracked the superhump wave of the black-hole X-ray transient KV Ursae Majoris (= XTE J1118+480) through its 2000 outburst. The properties of the superhump were a little different — smaller, more long-lived, more stable in period, more sinusoidal in waveform — from those of the common superhumps of dwarf novae. Nevertheless, the resemblances are sufficient to warrant adopting the hypothesis of a common origin. We found ε=0.0047(7).

4. We discuss the important constraints set by the nondetection of superhumps in U Geminorum (especially) and IP Pegasi. Assuming that there is a critical ratio $q_{\rm crit}$ for superhump manufacture, we infer that $q_{\rm crit}$ does not exceed 0.38.

5. We report a season's photometry on the novalike variable BB Doradus (= EC 05287–5857), yielding three noncommensurate frequencies. No help is available to guide us in interpreting the frequencies; but the strongest signal, at 6.701(3) c/d, can be plausibly associated with the orbital frequency $\omega_o$. We interpret the other two signals as the $\omega_o-\Omega$ and $2\omega_o-\Omega$ superhumps. This superhump is of special interest since it has a very large period excess ε=0.0939(15), thus providing a calibrating point (a limit) at large ε.

[28] They are *somewhat* larger than ZAMS stars, but this is probably due to the extra heating CV secondaries suffer as they struggle with thermal equilibrium (Paczynski & Sienkiewicz 1983). There is still room for abundance anomalies (C/N, etc.), but their interpretations should not invoke too much H burning.

6. DW Ursae Majoris supplies an additional calibration at large ε, now that eclipse analysis has supplied a *q* constraint.

7. We report survey results for ~200 stars, and tackle the question: which stars superhump? The answer is pretty simple, and not a property of dwarf-nova eruptions. Basically all short-period CVs which reach high-$\dot{M}$ states for more than a few days grow superhumps. As $P_{orb}$ increases from 2.3 hr to 3.5 hr, the superhumping fraction seems to drop smoothly from ~100% to ~15% — with the 50% threshold crossed at $P_{orb}$=3.1±0.2 hr. This smooth dependence on $P_{orb}$ is consistent with the hypothesis that the controlling parameter is *q*, with the width of the transition probably reflecting a variance in white-dwarf mass. We show how $q_{crit}$, $<M_1>$, and the secondary's mass–radius relation are linked.

8. Since superhump properties appear to be tightly correlated with *q* (independently measurable in ~12–15 cases) and $P_{orb}$ (measurable in all cases), we adopt the working hypothesis that *q* is the controlling parameter, and derive an estimate for $q_{crit}$. Superhumps are absent in all CVs with a certifiable $q$>0.36, and present in all CVs with a certifiable $q$<0.25. DW UMa and UU Aqr are of particular interest since they have a fairly high *q* (near 0.3) and yet show well-defined superhumps with ε not quite at the top of the range. We use this to estimate $q_{crit}$=0.35±0.02. This is also consistent with an estimate based on the 50% threshold at 3.1 hr.

9. We use all available data to establish an empirical ε(*q*) law, Eq. (8). This should be useful in estimating system parameters of any star showing superhumps. It also provides a mass–radius law for the secondary stars in short-period CVs, subject to the adopted value of $<M_1>$. Plausible choices are $<M_1>$=0.75 $M_\odot$, $R_2$=1.15 $R_{ZAMS}$. The resultant mass–radius law shows an apparent discontinuity near 0.20 $M_\odot$, which agrees with the expectation from a disrupted magnetic braking model.

10. The mass–radius law, shown in Figures 10–12 and Eq. (14), demonstrates that CV secondaries follow a theoretical ZAMS pretty well, with just two clear departures: near 0.08 and 0.20 $M_\odot$. The first is well-understood as the result of the star's inability to contract fast enough to keep up with its timescale of mass loss (first described by Paczynski 1981). The second is less understood, but probably arises from the same effect, where the timescale is set by the strength of magnetic braking. It should be noted, however, that these departures are mainly *gradual*. Essentially all CVs have secondaries slightly above the single-star ZAMS, and this bloating seems to steadily increase as they approach the transitions at 0.20 and 0.08 $M_\odot$. It's possible that this arises because the secondaries are always losing matter on timescales pretty close to their thermal timescales [see Figure 23 of Patterson (1984) and the accompanying discussion].

11. The very low dispersion in the $R_2(M_2)$ law is a striking result here — as if the evolving stars just slid gracefully down a well-defined string in Figures 10–12. These figures are somewhat misleading, though, because the uncertainty in $M_1$ moves the points roughly parallel to the apparent evolution tracks. So while low dispersion is true and is important, this is only true for the *collection* of stars; $M_2$ and $R_2$ are not so well determined for each individual star.

12. There is also a minor (at least) paradox in the understanding of a well-defined $R_2(M_2)$ law for short-period CVs. These stars range in apparent $<\dot{M}>$ from $10^{-11}$ $M_{\odot}$/yr (WZ Sge stars) to $10^{-9}$ (ER UMa stars) to $10^{-8}$ (CP Pup, BK Lyn). It is not easy to understand how the secondaries can muster a common mass–radius law in the face of such variety. Long-term mass-transfer cycles, possibly associated with classical nova eruptions, could perhaps explain this… but at present it looms as a mystery.

13. There are still some loose ends needing clarification. The question of U Gem superhumps is important and will affect our results. If the Smak & Waagen result is correct, then ε=0.130(14) at $q$=0.36(2). Adding this point to Figure 9 and refitting, we find a somewhat steeper ε($q$) with a larger quadratic term. That implies a smaller $q$ above the period gap, which would increase the discontinuity in radius we find near 0.2 $M_{\odot}$. TV Col needs clarification too, although cannot be easily compared with these other stars, since it lacks a $q$ constraint and is magnetic. And a radial-velocity study of BB Dor would be very welcome, since we found an annoying ambiguity of interpretation there. Finally, ε($q$) still needs help at the low-$q$ end; this will be critical in measuring $q$ and $M_2$ for the very oldest CVs, after they have passed minimum period.

This paper reports results from 262 nights and ~1000 hours — and that's just for the five stars with newly reported positive results. The negative results, and summary reports on stars not yet published, span another thousand nights. Keeping this collaboration humming along takes a lot of community mojo. Even outside the big author list, we benefited from the data of Gordon Garradd, Stan Walker, Bill Allen, Panos Niarchos, Bernard Heathcote, David Messier, Sarah Tuttle, Donn Starkey, and Fred Velthuis. The NSF provided some mojo too, in financial support through grants AST 00–98254 and 04–06813.

TABLE 1
OU VIR OBSERVING LOG
(2003 MAY – JUNE)

| Telescope(s) | Observer(s) | Nights/Hours |
| --- | --- | --- |
| MDM 2.4 m, 1.3 m | J. Kemp | 12/46 |
| CBA–Uzbekistan 28 cm | T. Krajci | 8/43 |
| CBA–Nelson 35 cm | R. Rea | 7/34 |
| CBA–Utah 50 cm | J. Foote | 4/25 |
| CBA–Flagstaff 40 cm | R. Fried | 3/18 |
| CBA–Pretoria 30 cm | B. Monard | 3/14 |
| CBA–Belgium 35 cm | T. Vanmunster | 3/11 |
| University of Athens | P. Niarchos | 2/14 |
| CBA–Connecticut 25 cm | D. Messier | 1/7 |
| CBA–East 66 cm | D. Skillman | 1/5 |
| CBA–Mia Mia | B. Heathcote | 1/4 |

TABLE 2
XZ ERI OBSERVING LOG
(2003 JANUARY – FEBRUARY)

| Telescope(s) | Observer(s) | Nights/Hours |
|---|---|---|
| MDM 1.3 m | E. Armstrong | 9/23 |
| CBA–Pretoria 30 cm | B. Monard | 6/22 |
| CBA–Perth 30 cm | G. Bolt | 6/22 |
| CBA–Nelson 35 cm | R. Rea | 5/16 |
| University of Auckland 35 cm | P. Warhurst | 3/5 |
| CBA–Blenheim 40 cm | B. Allen | 2/5 |
| CBA–Indiana 25 cm | D. Starkey | 1/5 |
| CBA–Utah 50 cm | J. Foote | 1/4 |

TABLE 3
UU AQR OBSERVING LOG
(2000)

| Telescope(s) | Observer(s) | Nights/Hours |
|---|---|---|
| CBA–Flagstaff 40 cm | R. Fried | 11/62 |
| CBA–Farm Cove 25 cm | J. McCormick, F. Velthuis | 9/39 |
| CBA–Denmark 25 cm | L. Jensen | 8/26 |
| CBA–Illinois 20 cm | J. Gunn | 7/24 |
| CBA–Nelson 35 cm | R. Rea | 6/32 |
| CBA–Townsville 20 cm | N. Butterworth | 6/26 |
| CBA–Otahuhu 30 cm | M. Bos | 4/24 |
| CBA–Concord 44 cm | L. Cook | 2/10 |
| CBA–Tamworth 45 cm | G. Garradd | 2/9 |
| CBA–Waiharara 25 cm | S. Walker | 2/9 |
| CBA–Belgium 25 cm | T. Vanmunster | 2/7 |
| CTIO 0.9 m | J. Kemp | 1/2 |

TABLE 4
KV UMA (= XTE J1118+480) OBSERVING LOG

| Telescope(s) | Observer(s) | Nights/Hours |
|---|---|---|
| CBA–West 35 cm | D. Harvey | 32/180 |
| CBA–Concord 44 cm | L. Cook | 18/63 |
| CBA–East 66 cm | D. Skillman | 16/51 |
| CBA–Flagstaff 40 cm | R. Fried | 14/101 |
| CBA–Belgium 25 cm | T. Vanmunster | 10/46 |
| MDM 1.3 m, 2.4 m | J. Kemp | 10/15 |
| CBA–Italy 28 cm | G. Masi | 3/11 |
| Interval (JD) | Superhump Period (d) | Semi-Amplitude (mag) |
| 2451634–48 | 0.17094(20) | 0.028 |
| 2451647–60 | 0.17074(14) | 0.029 |
| 2451657–71 | 0.17027(27) | 0.037 |
| 2451670–83 | 0.17047(22) | 0.034 |
| 2451686–99 | 0.17065(22) | 0.036 |
| 2451700–10 | 0.17062(26) | 0.040 |
| Date (JD) | QPO Period (s) | |
| 2451658 | 11.5±0.2 | |
| 2451665 | 9.9±0.1 | |
| 2451677 | 8.8±0.2 | |

TABLE 5
BB DOR (= EC 05287–5857)

| Telescope(s) | Observer(s) | Nights/Hours |
|---|---|---|
| CBA–Pretoria 30 cm | B. Monard | 13/65 |
| CTIO 0.9 m | J. Kemp | 10/53 |
| CBA–Nelson 35 cm | R. Rea | 4/10 |
| CBA–Perth 30 cm | G. Bolt | 3/12 |

TABLE 6
SUPERHUMP SUCCESS RATE

| $P_{orb}$ (d) | Superhumps/Searched |
|---|---|
| 0.05→0.06 | 22/24 |
| 0.06→0.07 | 40/41 |
| 0.07→0.08 | 37/38 |
| 0.08→0.10 | 20/23 |
| 0.10→0.13 | 9/13 |
| 0.13→0.15 | 8/25 |
| 0.15→0.17 | 2/8 |
| 0.17→0.20 | 0/11 |
| 0.20→0.30 | 0/13 |
| 0.30→0.40 | 0/4 |
| 0.40→0.50 | 0/4 |
| 0.50→0.60 | 0/2 |

NOTE — These refer to stars with detections and strong upper limits (roughly <0.04 mag). Weak upper limits (~0.1 mag) are not counted as searches.

TABLE 7
SUPERHUMP FRACTIONAL PERIOD EXCESS VERSUS MASS RATIO

| Star | ε | *q* | References |
|---|---|---|---|
| KV UMa (= XTE J1118+480) | 0.0047(9) | 0.037(7) | this paper, Uemura et al. 2002, Orosz 2001, Zurita et al. 2002 |
| WZ Sge | 0.0092(8) | 0.050(15) | Patterson et al. 2002a, Steeghs et al. 2001 |
| XZ Eri | 0.0270(15) | 0.110(2) | this paper, Feline et al. 2004b, Uemura et al. 2004 |
| IY UMa | 0.0260(10) | 0.125(8) | Patterson et al. 2000c, Steeghs et al. 2003 |
| Z Cha | 0.0364(9) | 0.145(15) | Warner & O'Donoghue 1988, Wade & Horne 1988 |
| HT Cas | 0.0330(30) | 0.15(1) | Zhang et al. 1986, Horne et al. 1991 |
| DV UMa | 0.0343(10) | 0.150(1) | Patterson et al. 2000b, Feline et al. 2004b |
| OU Vir | 0.0326(15) | 0.175(25) | this paper, Feline et al. 2004a |
| V2051 Oph | 0.030(2) | 0.19(3) | Baptista et al. 1998, Kiyota & Kato 1998, P03 |
| DW UMa | 0.0644(20) | 0.28(4) | Araujo-Betancor et al. 2003, this paper, Patterson et al. 2002b |
| UU Aqr | 0.0702(19) | 0.30(7) | Baptista et al. 1994, this paper |
| BB Dor (= EC 05287–5857) | 0.0939(15) | <0.38 | this paper |

NOTES —

1. We generally list errors from the published source, but include a (usually) larger error of 5% in both parameters in the fit, and in Figure 10. Such an error is always appropriate for ε, because superhumps commonly show small period changes. Much smaller errors in *q* are sometimes quoted in the literature; but these are always internal errors, not including the (usually unknown) systematic error in the method used.
2. In P01 we included X-ray binaries in this discussion. They are somewhat relevant, and do as a class establish that low ε is always associated with low *q* (compare our Figure 9 to Figure 1 of P01). But the observational errors in these stars are generally too large to be a significant constraint, and there are some worries about cycle count too; so we omit them here. The one exception is KV UMa (= XTE J1118+480).

TABLE 8
BASIC DATA ON ECLIPSING CATACLYSMIC VARIABLES
($P_{\rm ORB}$ < 8 HR)

| Star | $M_1$ ($M_\odot$) | $M_2$ ($M_\odot$) | $R_2$ ($R_\odot$) | $P_{\rm orb}$ (d) | $q$ | References |
|---|---|---|---|---|---|---|
| XZ Eri | 0.77(2) | 0.084(2) | 0.131(2) | 0.06116 | 0.110(2) | Feline et al. 2004b |
| OY Car | 0.685(11) | 0.070(2) | 0.127(2) | 0.06312 | 0.10(1) | Wood et al. 1989 |
| EX Hya | 0.49(13) | 0.081(13) | 0.138(6) | 0.06823 | 0.165(30) | Hoogerwerf et al. 2004, Beuermann et al. 2003 |
| OU Vir | 0.9(2) | 0.15(4) | 0.177(24) | 0.07271 | 0.175(25) | Feline et al. 2004a |
| HT Cas | 0.61(4) | 0.09(2) | 0.154(13) | 0.07365 | 0.15(1) | Horne et al. 1991 |
| IY UMa | 0.79(4) | 0.10(1) | 0.160(4) | 0.07391 | 0.125(8) | Steeghs et al. 2003 |
| Z Cha | 0.56(1) | 0.083(3) | 0.149(4) | 0.0745 | 0.145(15) | Wade & Horne 1988, Wood 1990, Wood et al. 1986 |
| DV UMa | 1.09(8) | 0.15(1) | 0.207(16) | 0.08585 | 0.151(1) | Feline et al. 2004b |
| DW UMa | 0.73(3) | 0.21(3) | 0.305(15) | 0.13661 | 0.28(4) | Araujo-Betancor et al. 2003, this paper |
| IP Peg | 0.94(10) | 0.42(8) | 0.43(3) | 0.15821 | 0.45(4) | Smak 2002, Martin et al. 1989, Hessman 1989 |
| UU Aqr | 0.67(14) | 0.20(7) | 0.34(4) | 0.16358 | 0.30(7) | Baptista et al. 1994 |
| GY Cnc | 0.82(14) | 0.33(7) | 0.41(3) | 0.17544 | 0.40(8) | Thorstensen 2000 |
| U Gem | 1.07(8) | 0.39(2) | 0.45(1) | 0.17691 | 0.36(2) | Smak 2001 |
| DQ Her | 0.60(7) | 0.40(5) | 0.48(3) | 0.19362 | 0.66(4) | Horne et al. 1993 |
| EX Dra | 0.75(15) | 0.55(8) | 0.58(4) | 0.20994 | 0.72(6) | Baptista et al. 2000 |
| V347 Pup | 0.63(4) | 0.52(6) | 0.60(2) | 0.23194 | 0.83(5) | Thoroughgood et al. 2005 |
| EM Cyg | 1.12(8) | 0.99(12) | 0.87(7) | 0.29091 | 0.88(5) | North et al. 2000 |
| AC Cnc | 0.76(3) | 0.77(5) | 0.83(3) | 0.30048 | 1.02(4) | Thoroughgood et al. 2004 |
| V363 Aur | 0.90(6) | 1.06(11) | 0.97(4) | 0.32124 | 1.17(7) | Thoroughgood et al. 2004 |

TABLE 9
$M_2$ AND $R_2$ FROM SUPERHUMPS

| Star | $P_{orb}$ (d) | ε | $q$ | $R_2$ ($R_\odot$) | $M_2$ ($M_\odot$) | $M_1$** ($M_\odot$) |
|---|---|---|---|---|---|---|
| DI UMa | 0.05456(1) | 0.0133(6) | 0.070(4) | 0.105(5) | 0.053(3) | – |
| V844 Her | 0.05464(1) | 0.0243(9) | 0.115(5) | 0.125(4) | 0.086(4) | – |
| LL And | 0.05505(1) | 0.0290(36) | 0.131(15) | 0.131(6) | 0.098(11) | – |
| SDSS 0137–09 | 0.05537(4) | 0.0248(20) | 0.116(10) | 0.126(6) | 0.088(7) | – |
| ASAS 0025+12 | 0.05605(5)* | 0.0206(21) | 0.096(9) | 0.119(4) | 0.072(7) | – |
| AL Com | 0.05667(3) | 0.0120(7) | 0.060(4) | 0.103(4) | 0.045(3) | – |
| WZ Sge | 0.05669(1) | 0.0092(7) | 0.046(4) | 0.107(3) | 0.051(4) | 1.1 |
| RX 1839+26 | 0.05669(5)* | 0.0173(20) | 0.083(10) | 0.115(5) | 0.063(7) | – |
| PU CMa | 0.05669(5) | 0.0222(20) | 0.109(11) | 0.122(5) | 0.083(9) | – |
| SW UMa | 0.05681(14) | 0.0245(27) | 0.113(13) | 0.127(6) | 0.085(9) | – |
| HV Vir | 0.05707(1) | 0.0200(9) | 0.094(6) | 0.120(3) | 0.071(5) | – |
| MM Hya | 0.05759(1) | 0.0184(10) | 0.086(6) | 0.117(3) | 0.065(4) | – |
| WX Cet | 0.05829(4) | 0.0199(15) | 0.094(9) | 0.122(5) | 0.071(7) | – |
| KV Dra | 0.05876(7) | 0.0233(22) | 0.107(11) | 0.128(6) | 0.080(8) | – |
| T Leo | 0.05882(1) | 0.0236(14) | 0.110(8) | 0.129(4) | 0.083(8) | – |
| EG Cnc | 0.05997(9) | 0.0067(8) | 0.035(6) | 0.089(6) | 0.027(5) | – |
| V1040 Cen | 0.06028(10) | 0.0310(27) | 0.139(13) | 0.142(5) | 0.104(10) | – |
| RX Vol | 0.0603(2)* | 0.0178(20) | 0.086(11) | 0.121(6) | 0.065(7) | – |
| AQ Eri | 0.06094(6) | 0.0284(21) | 0.129(11) | 0.139(5) | 0.097(9) | – |
| XZ Eri | 0.06116(1) | 0.0270(16) | 0.123(7) | 0.138(3) | 0.094(5) | 0.77 |
| CP Pup | 0.06145(6) | 0.0171(20) | 0.082(11) | 0.132(5) | 0.082(7) | 1.0 |
| V1159 Ori | 0.06218(1) | 0.0320(11) | 0.142(6) | 0.146(3) | 0.107(4) | – |
| V2051 Oph | 0.06243(1) | 0.0281(25) | 0.127(13) | 0.141(5) | 0.095(10) | – |
| V436 Cen | 0.0625(2) | 0.0212(32) | 0.100(16) | 0.130(9) | 0.075(12) | – |
| BC UMa | 0.06261(1) | 0.0306(14) | 0.137(8) | 0.145(4) | 0.103(6) | – |
| HO Del | 0.06266(16) | 0.0276(35) | 0.125(17) | 0.140(8) | 0.094(12) | – |
| EK TrA | 0.06288(5) | 0.0321(25) | 0.142(13) | 0.147(5) | 0.107(10) | – |
| TV Crv | 0.0629(2) | 0.0325(32) | 0.144(14) | 0.148(6) | 0.108(10) | – |
| VY Aqr | 0.06309(4) | 0.0203(15) | 0.095(7) | 0.129(5) | 0.071(6) | – |
| OY Car | 0.06312(1) | 0.0203(15) | 0.095(7) | 0.126(5) | 0.069(6) | 0.69 |
| RX 1131+43 | 0.06331(8) | 0.0259(16) | 0.119(9) | 0.139(5) | 0.089(7) | – |
| ER UMa | 0.06366(3) | 0.0314(11) | 0.140(5) | 0.147(4) | 0.105(4) | – |
| DM Lyr | 0.06546(6) | 0.0281(31) | 0.127(14) | 0.145(6) | 0.095(10) | – |
| UV Per | 0.06489(11) | 0.0234(23) | 0.108(10) | 0.137(5) | 0.081(8) | – |
| AK Cnc | 0.0651(2) | 0.0368(33) | 0.160(16) | 0.156(7) | 0.120(12) | – |
| AO Oct | 0.06557(13) | 0.0242(39) | 0.111(17) | 0.139(7) | 0.083(12) | – |
| SX LMi | 0.06717(11) | 0.0347(25) | 0.153(11) | 0.157(5) | 0.115(8) | – |
| SS UMi | 0.06778(4) | 0.0360(15) | 0.158(12) | 0.160(5) | 0.119(9) | – |
| KS UMa | 0.06796(10) | 0.0241(30) | 0.112(13) | 0.143(6) | 0.084(10) | – |
| V1208 Tau | 0.0681(2) | 0.0374(28) | 0.163(15) | 0.162(6) | 0.122(12) | – |
| RZ Sge | 0.06828(2) | 0.0306(28) | 0.137(13) | 0.153(5) | 0.103(10) | – |
| TY Psc | 0.06833(5) | 0.0347(15) | 0.153(12) | 0.159(5) | 0.115(9) | – |
| IR Gem | 0.0684(3) | 0.0351(66) | 0.154(30) | 0.160(10) | 0.115(9) | – |
| V699 Oph | 0.0689(2)* | 0.0197(28) | 0.093(11) | 0.136(8) | 0.070(9) | – |
| CY UMa | 0.06957(4) | 0.0364(14) | 0.159(10) | 0.163(4) | 0.119(7) | – |
| FO And | 0.07161(18) | 0.0349(40) | 0.153(17) | 0.164(7) | 0.115(12) | – |
| OU Vir | 0.07271(1) | 0.0326(15) | 0.145(7) | 0.184(4) | 0.156(7) | 0.90 |
| VZ Pyx | 0.07332(3) | 0.0333(20) | 0.147(9) | 0.163(5) | 0.110(7) | – |

(CONTINUED ON NEXT PAGE)

(CONTINUED FROM PREVIOUS PAGE)

TABLE 9
$M_2$ AND $R_2$ FROM SUPERHUMPS

| Star | $P_{orb}$ (d) | ε | $q$ | $R_2$ ($R_\odot$) | $M_2$ ($M_\odot$) | $M_1$** ($M_\odot$) |
|---|---|---|---|---|---|---|
| CC Cnc | 0.07352(5) | 0.0487(27) | 0.203(12) | 0.184(5) | 0.152(9) | – |
| HT Cas | 0.07365(1) | 0.0330(30) | 0.147(12) | 0.154(5) | 0.090(9) | 0.61 |
| IY UMa | 0.07391(1) | 0.0260(13) | 0.120(7) | 0.155(4) | 0.091(5) | 0.79 |
| VW Hyi | 0.07427(1) | 0.0331(8) | 0.147(5) | 0.166(4) | 0.110(4) | – |
| Z Cha | 0.07450(1) | 0.0364(9) | 0.159(6) | 0.155(3) | 0.090(4) | 0.56 |
| QW Ser | 0.07453(10) | 0.0331(40) | 0.147(18) | 0.166(7) | 0.110(12) | – |
| WX Hyi | 0.07481(1) | 0.0346(14) | 0.152(11) | 0.169(5) | 0.114(9) | – |
| BK Lyn | 0.07498(5) | 0.0479(7) | 0.200(7) | 0.185(4) | 0.150(5) | – |
| RZ Leo | 0.07604(1) | 0.0347(25) | 0.152(13) | 0.170(6) | 0.114(10) | – |
| AW Gem | 0.07621(10) | 0.0422(27) | 0.180(18) | 0.181(7) | 0.135(12) | – |
| SU UMa | 0.07635(5) | 0.0317(12) | 0.141(10) | 0.167(5) | 0.106(7) | – |
| SDSS 1730+62 | 0.07655(9) | 0.0376(22) | 0.162(11) | 0.175(6) | 0.122(8) | – |
| HS Vir | 0.0769(2) | 0.0477(23) | 0.193(15) | 0.186(6) | 0.145(10) | – |
| V503 Cyg | 0.0777(2) | 0.0430(27) | 0.183(13) | 0.184(5) | 0.138(9) | – |
| V359 Cen | 0.0779(3) | 0.0388(40) | 0.168(17) | 0.179(8) | 0.126(11) | – |
| CU Vel | 0.0785(2) | 0.0293(36) | 0.131(20) | 0.161(8) | 0.098(13) | – |
| NSV 9923 | 0.0791(2)* | 0.0412(30) | 0.175(12) | 0.183(6) | 0.131(9) | – |
| BR Lup | 0.0795(2) | 0.0340(40) | 0.150(18) | 0.175(8) | 0.113(13) | – |
| V1974 Cyg | 0.08126(1) | 0.0471(10) | 0.197(5) | 0.194(3) | 0.148(4) | 1.0 |
| TU Crt | 0.08209(9) | 0.0397(22) | 0.170(11) | 0.186(6) | 0.128(8) | – |
| TY PsA | 0.08414(18) | 0.0417(22) | 0.178(11) | 0.192(5) | 0.134(7) | – |
| KK Tel | 0.08453(21) | 0.0368(31) | 0.160(17) | 0.187(8) | 0.120(11) | – |
| V452 Cas | 0.08460(20) | 0.0497(33) | 0.206(17) | 0.203(9) | 0.155(12) | – |
| DV UMa | 0.08585(1) | 0.0343(11) | 0.151(6) | 0.208(4) | 0.163(4) | 1.09 |
| YZ Cnc | 0.0868(2) | 0.0553(26) | 0.224(12) | 0.212(6) | 0.168(9) | – |
| GX Cas | 0.08902(16) | 0.0449(25) | 0.190(11) | 0.204(6) | 0.143(7) | – |
| NY Ser | 0.09775(19) | 0.0623(35) | 0.247(16) | 0.237(7) | 0.185(12) | – |
| V348 Pup | 0.10184(1) | 0.0640(40) | 0.253(16) | 0.245(7) | 0.190(12) | – |
| V795 Her | 0.10826(1) | 0.0760(10) | 0.290(5) | 0.288(4) | 0.217(5) | – |
| V592 Cas | 0.11506(1) | 0.0625(5) | 0.248(5) | 0.265(4) | 0.186(4) | – |
| TU Men | 0.1172(2) | 0.0717(32) | 0.277(15) | 0.278(8) | 0.208(12) | – |
| AH Men | 0.12721(6) | 0.0887(16) | 0.326(10) | 0.310(6) | 0.244(8) | – |
| DW UMa | 0.13661(1) | 0.0644(20) | 0.255(10) | 0.300(6) | 0.191(8) | 0.73 |
| TT Ari | 0.13755(1) | 0.0847(7) | 0.315(7) | 0.323(5) | 0.236(6) | – |
| V603 Aql | 0.1381(2) | 0.0572(51) | 0.232(23) | 0.332(11) | 0.232(23) | 1.0 |
| PX And | 0.14635(1) | 0.0898(14) | 0.329(11) | 0.341(7) | 0.247(8) | – |
| V533 Her | 0.1473(2) | 0.0719(20) | 0.285(12) | 0.327(6) | 0.214(9) | – |
| BB Dor | 0.1492(1) | 0.0939(10) | 0.341(20) | 0.350(8) | 0.256(15) | – |
| BH Lyn | 0.15575(1) | 0.0790(30) | 0.301(15) | 0.346(8) | 0.226(15) | – |
| UU Aqr | 0.16358(1) | 0.0702(14) | 0.271(9) | 0.340(10) | 0.195(18) | 0.67 |

* $P_{orb}$ inferred from superhump sideband ($n\omega_{sh}+m\Omega$, where $m$=1, 2, ..., $n$) in superoutburst time series; these values are less well determined, and less reliable, than other measures of $P_{orb}$.

** $M_1$ assumed to be 0.75 $M_\odot$, except in a few listed cases where a better constraint is available. $M_2$ scales as $M_1/0.75\ M_\odot$, and $R_2$ scales as $(M_1/0.75\ M_\odot)^{1/3}$. We reckon that the main uncertainty in $M_2$ and $R_2$ arises from the $M_1$ dependence, rather than in errors of ε or the adopted ε($q$) relation.

## FIGURE CAPTIONS

FIGURE 1. — *Upper left frame*, eruption light curve of OU Vir. The open squares are upper limits. *Upper right*, the mean orbital waveform in superoutburst. *Middle frame*, 8-day light curve in superoutburst, after removing the mean, trend, and eclipses for each night. *Lower frame*, power spectrum of the 8-day light curve, with significant signals marked with their frequencies in c/d (±0.02). The mean superhump waveform is inset.

FIGURE 2. — *Upper frame*: sample light curve of XZ Eri in superoutburst. *Lower frame*: power spectrum of the 10-night spliced light curve, with significant frequencies marked in c/d. The observation window produced weak sidebands ±3.0 c/d displaced from the main peaks, which accounts for the odd appearance. Inset is the mean superhump light curve.

FIGURE 3. — Power spectrum of the UU Aqr time series in 2000, after removal of eclipses. Significant signals are marked with their frequency in c/d (±0.008). In order of increasing frequency, these signals are interpreted as $\omega_o-\Omega$, $2(\omega_o-\Omega)$, $3(\omega_o-\Omega)$, and $3\omega_o-2\Omega$.

FIGURE 4. — The power spectrum of the first 30 days of coverage of KV UMa (= XTE J1118+480). A simple signal at 5.857(7) c/d is seen, with a sinusoidal waveform (inset).

FIGURE 5. — O–C diagram of the timings of maximum light of KV UMa, relative to a test period of 0.17065 d. The fitted parabola indicates a slowly decreasing period, with $\dot{P}=-1\times10^{-6}$.

FIGURE 6. — *Left frame*, average power spectrum of 36 10-minute observations at 2 s time resolution on JD2451665. A quasi-period oscillation centered on $P$=9.9 s is evident. *Right frame*, other detections of the QPO.

FIGURE 7. — *Top frame*, average nightly power spectrum of BB Dor (= EC 05287–5857). *Middle frame*, power spectrum of the 45-night light curve, with significant signals marked with their frequencies in c/d (±0.003). *Bottom frame*, mean orbital (6.701 c/d) and superhump (6,126 c/d) waveforms.

FIGURE 8. — Percentage of (apsidal) superhumps as a function of $P_{\mathrm{orb}}$, drawn from Table 6. $P_{\mathrm{orb}}$ appears to be a good predictor, with 50% occurring at $P_{\mathrm{orb}}$=3.1±0.2 hrs. The width of the transition allows us to limit the dispersion in $M_1$, and in the mass–radius law.

FIGURE 9. — Variation of ε with $q$ in the stars of Table 7. The limit for BB Dor ($q$<0.38) is set by the requirement that it should have a lower $q$ than U Gem, since the latter (probably) does not superhump. The fitted quadratic curve is Eq. (8).

FIGURE 10. — Mass–radius dependence for secondary stars in CVs. Filled squares are ($M_2$, $R_2$) pairs deduced from superhumps. Diamonds show typical errors in the short-$P_{\mathrm{orb}}$ and long-$P_{\mathrm{orb}}$ regimes. Crosses are values independently deduced in eclipsing binaries (from Table 8). Enclosing boxes denote "long-period" stars (>2.5 hours). The curve shows the *M–R* dependence on the ZAMS (Baraffe et al. 1998), extended to lower mass with the brown dwarf models of Burrows et al. (1993). All CV secondaries are slightly too big to be ZAMS stars, but there is an apparent discontinuity near 0.20 $M_\odot$.

FIGURE 11. — Secondary-star radii compared to ZAMS radii, as a function of mass. This is obtained by binning the data of Figure 10 (0.01 and 0.02 $M_\odot$ bins). The general form of this curve is a secure result, unless the $\varepsilon(q)$ relation is greatly in error. But the location of the break at 0.19 $M_\odot$ depends on $< M_1 >$, which could be off by ~15%.

FIGURE 12. — Mass–radius plot in log–log space, extended to 1 $M_\odot$ and including additional eclipsing stars from Table 8. Straight lines are the empirical fits given by Eq. (14).

Fig 1

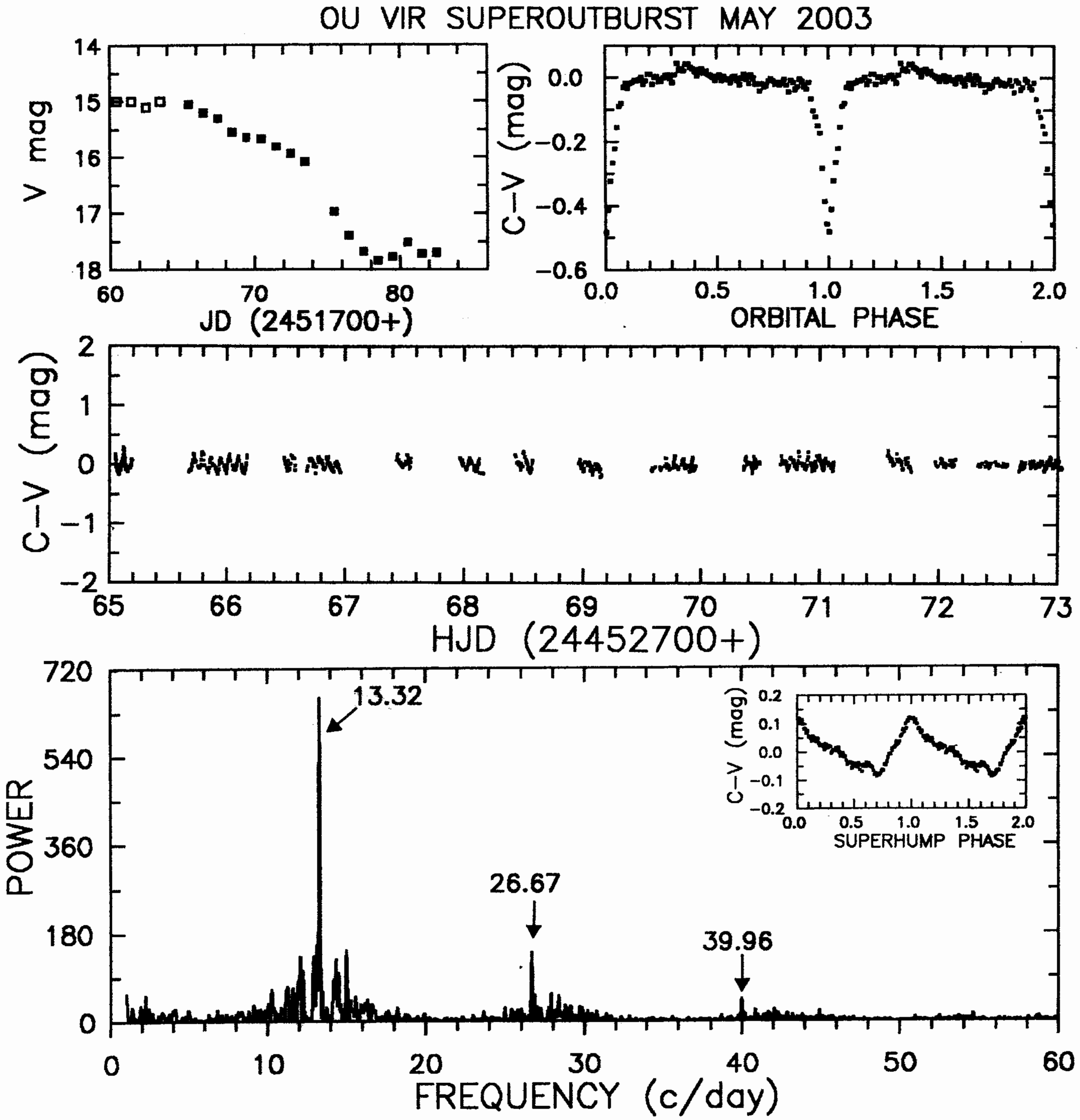

Fig 2

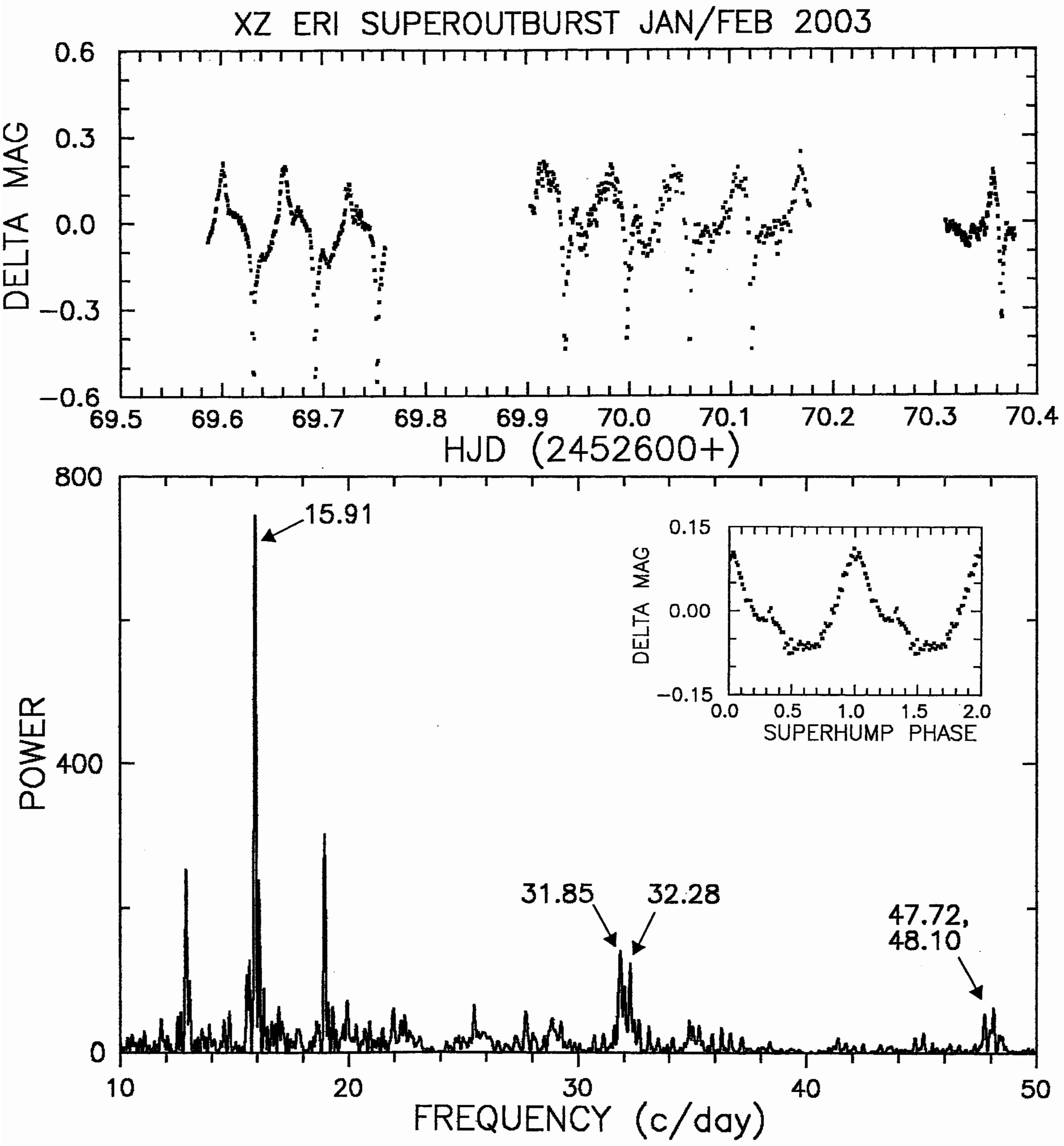

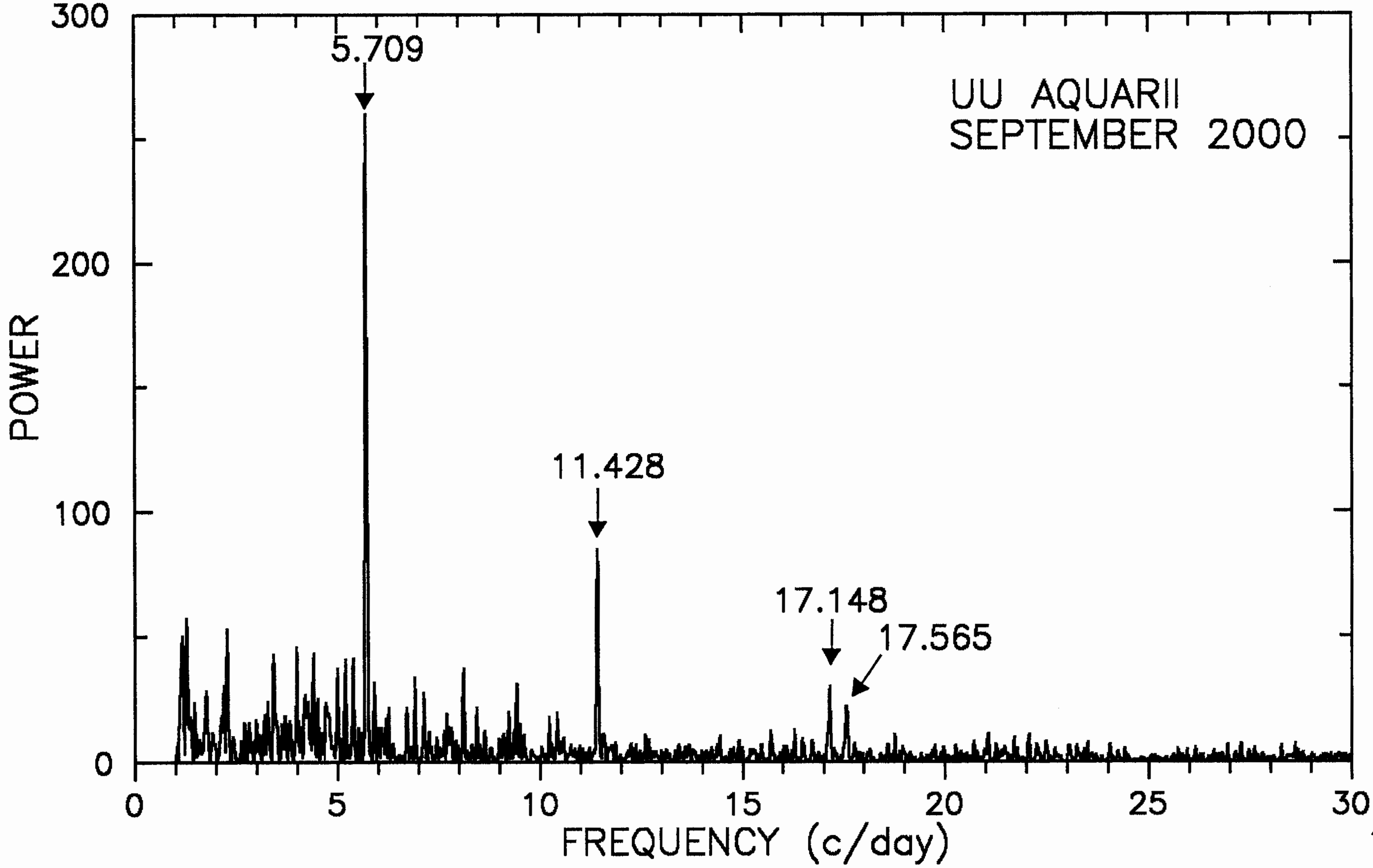

Fig 3

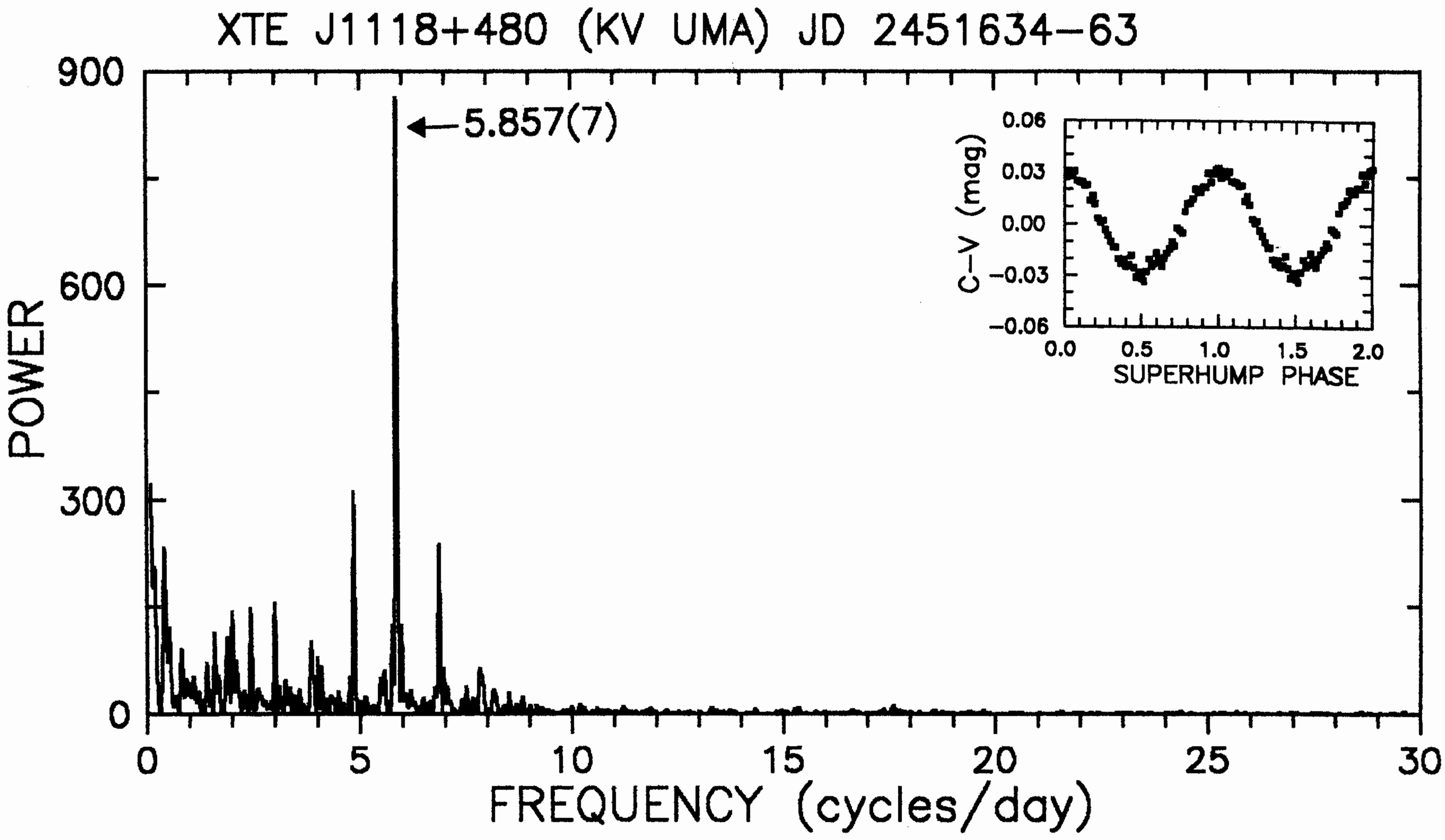

Fig 4

Fig 5

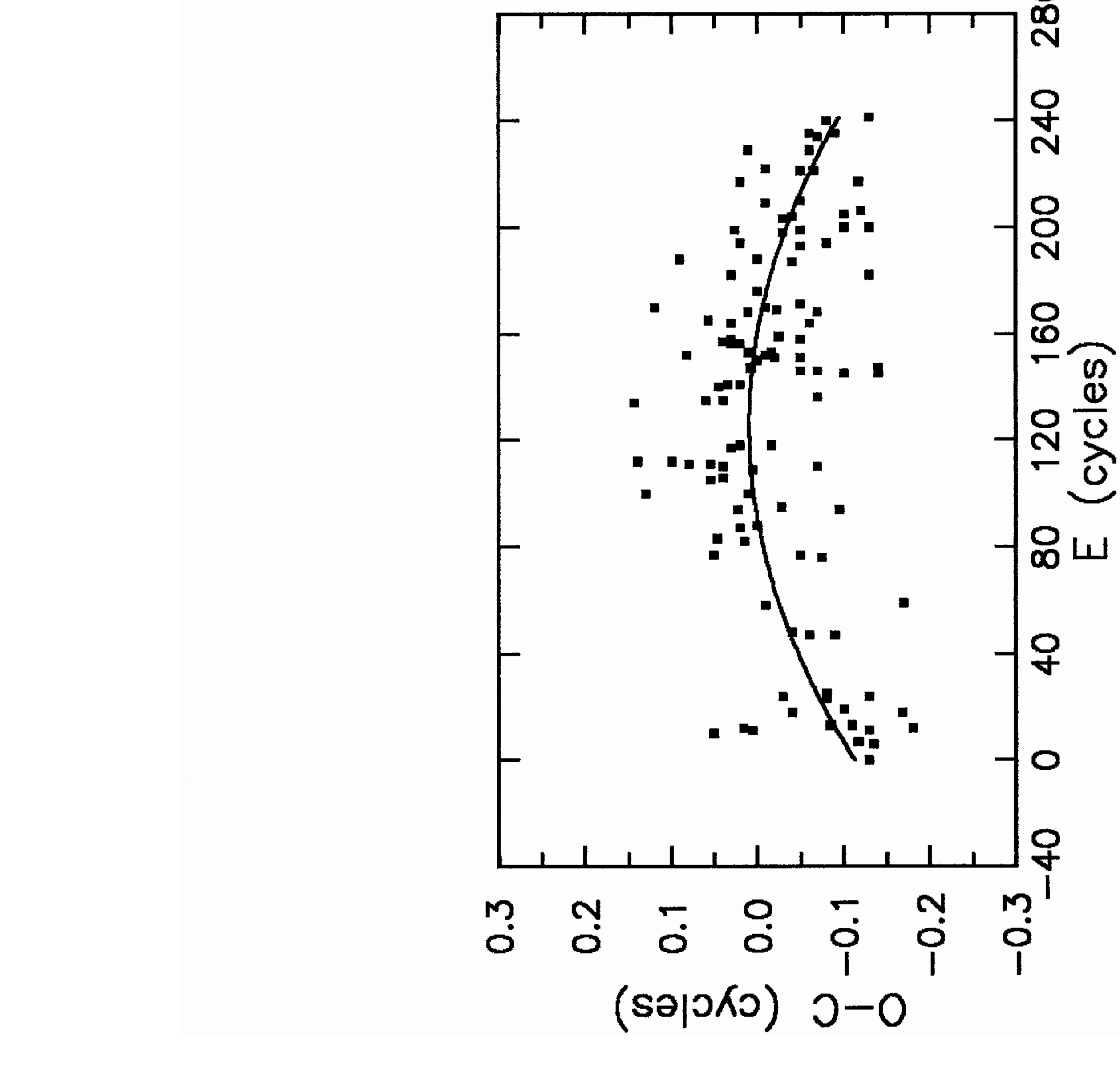

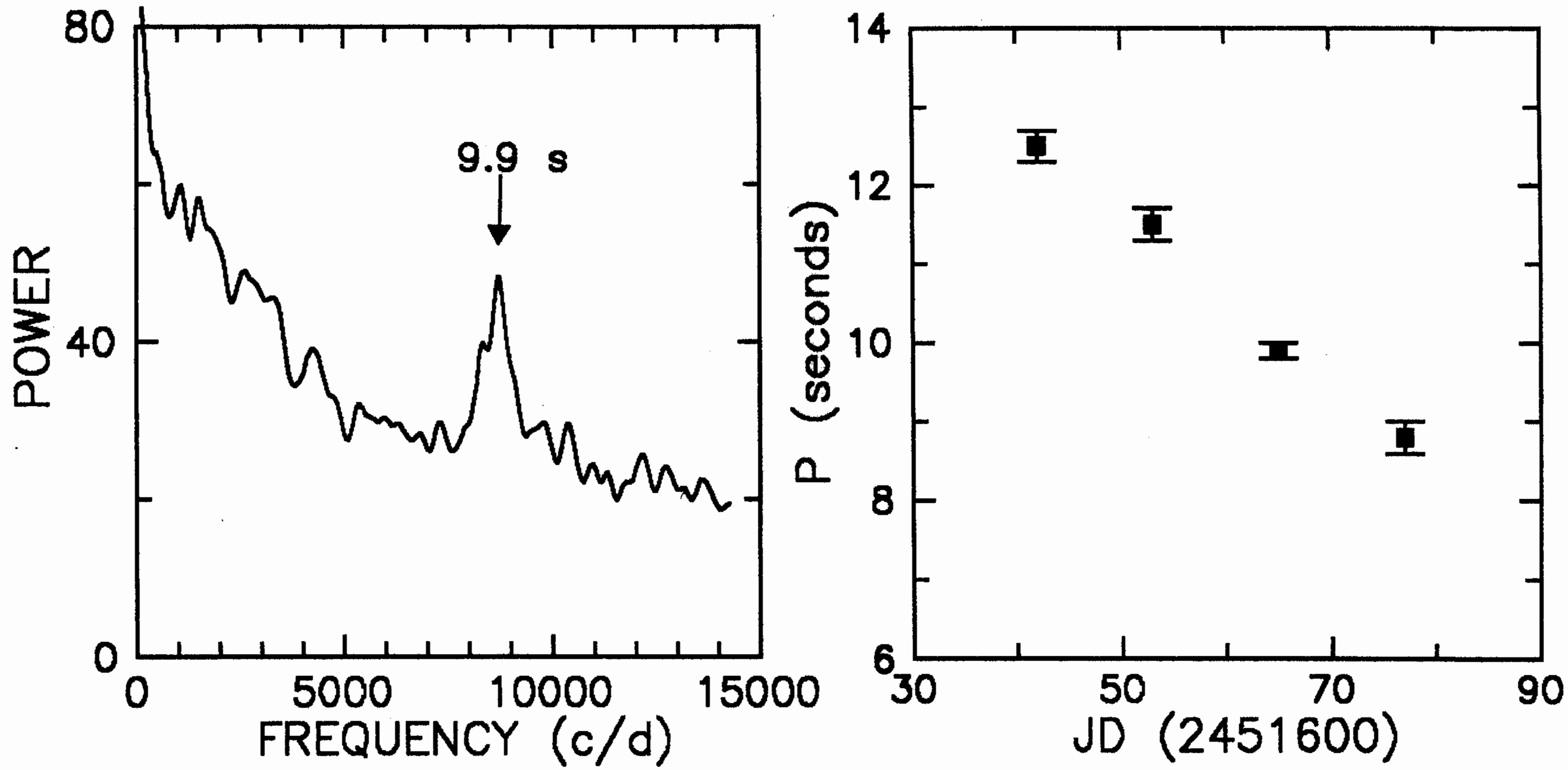

Fig 6

Fig 7

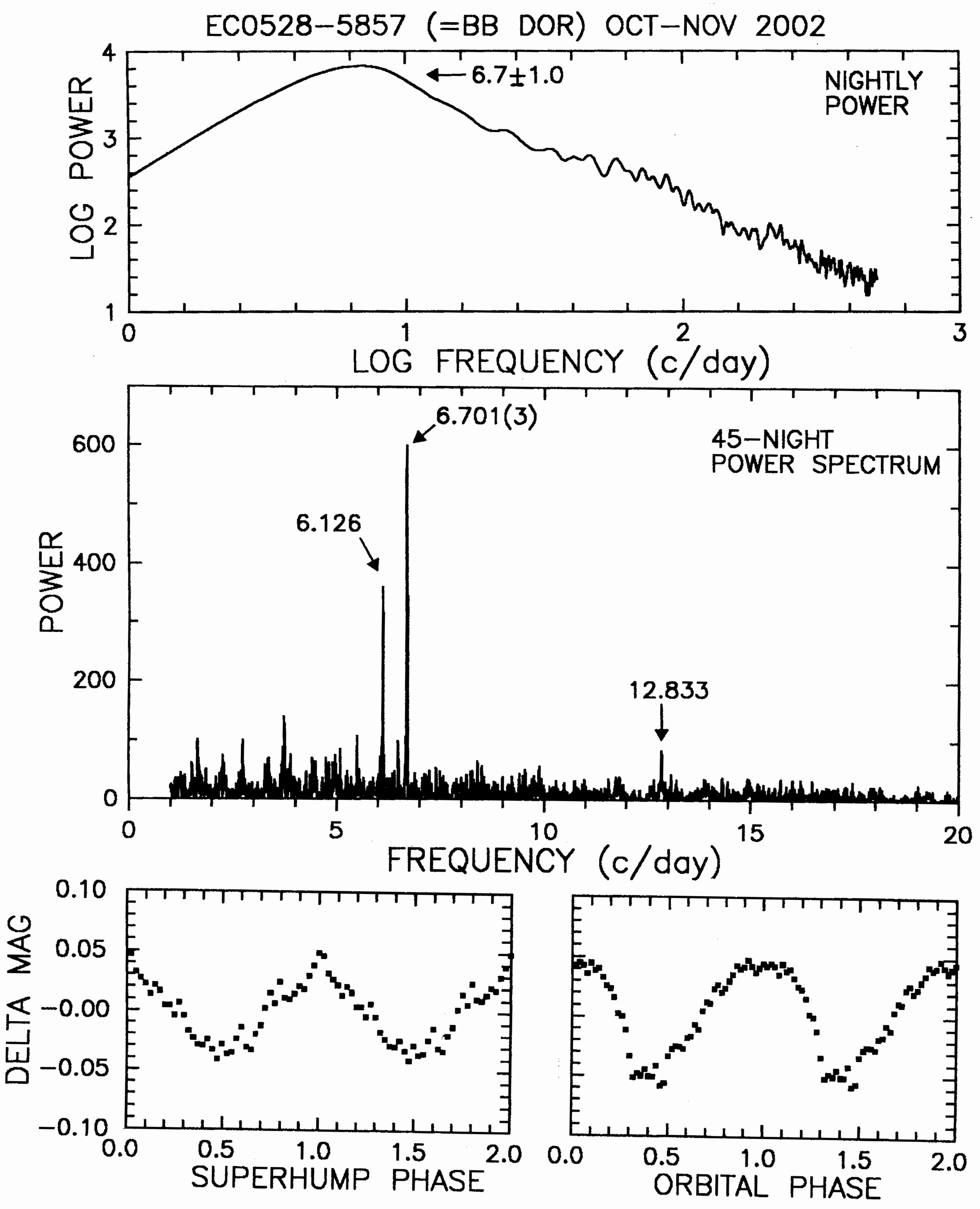

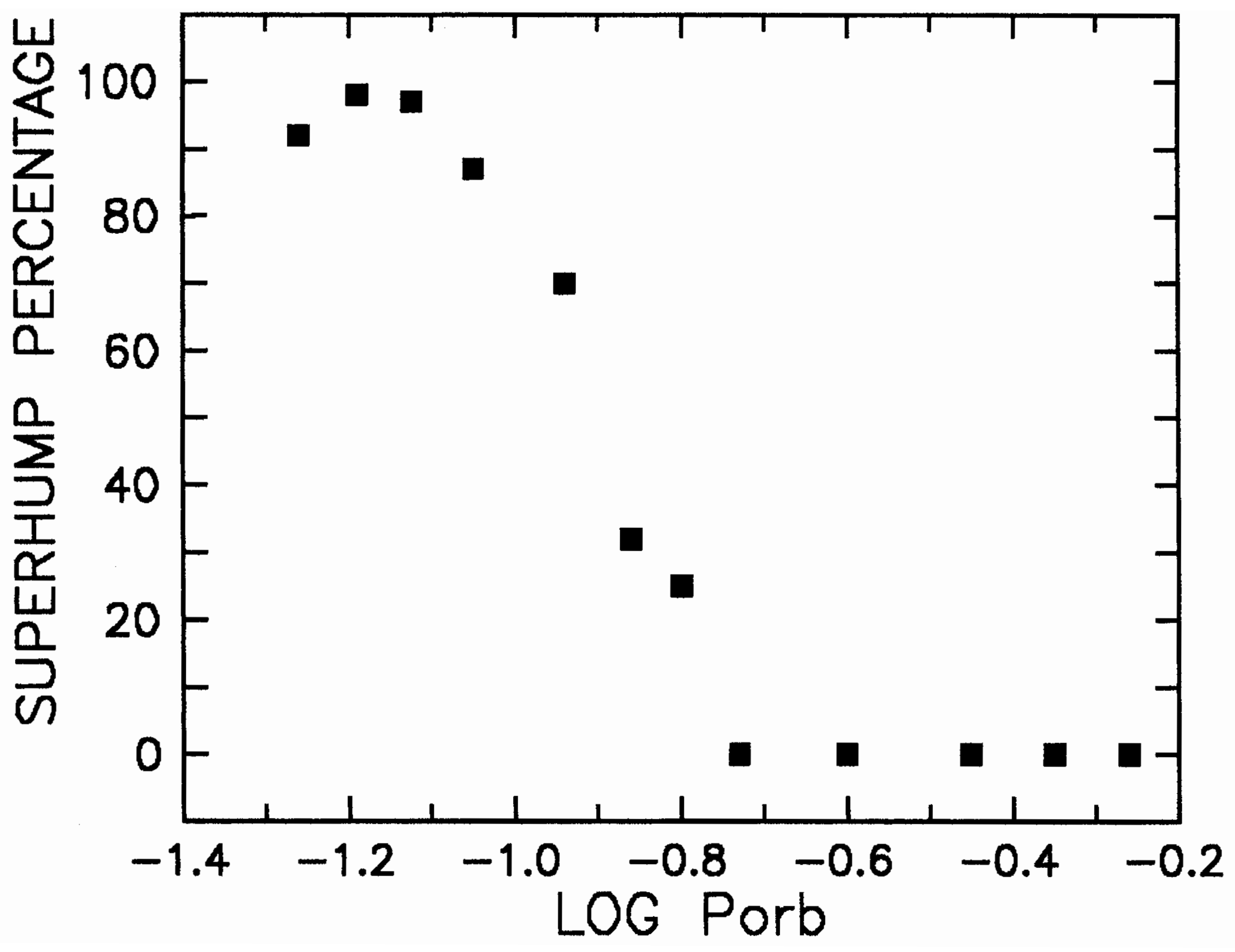

Fig 8

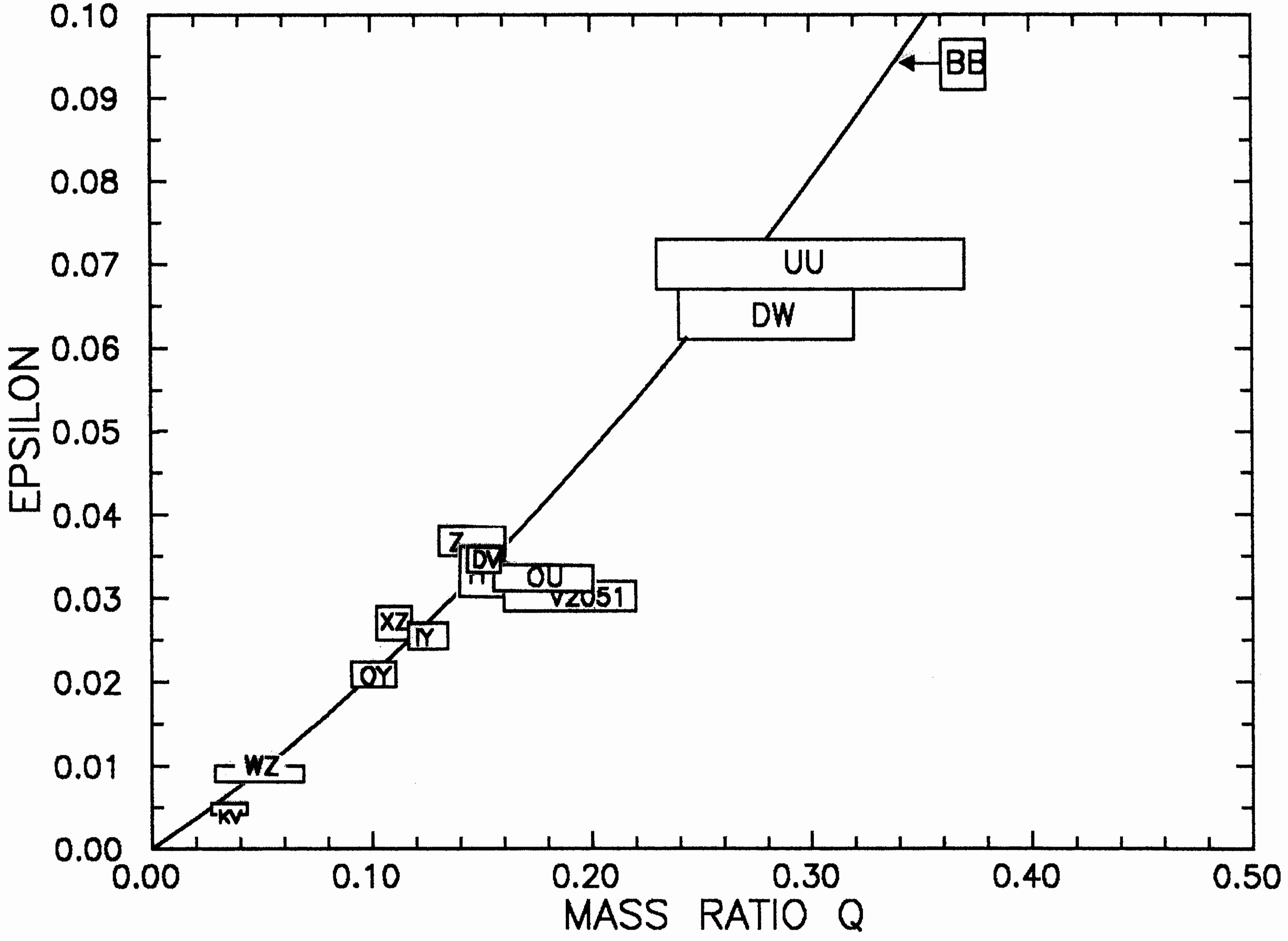

Fig 9

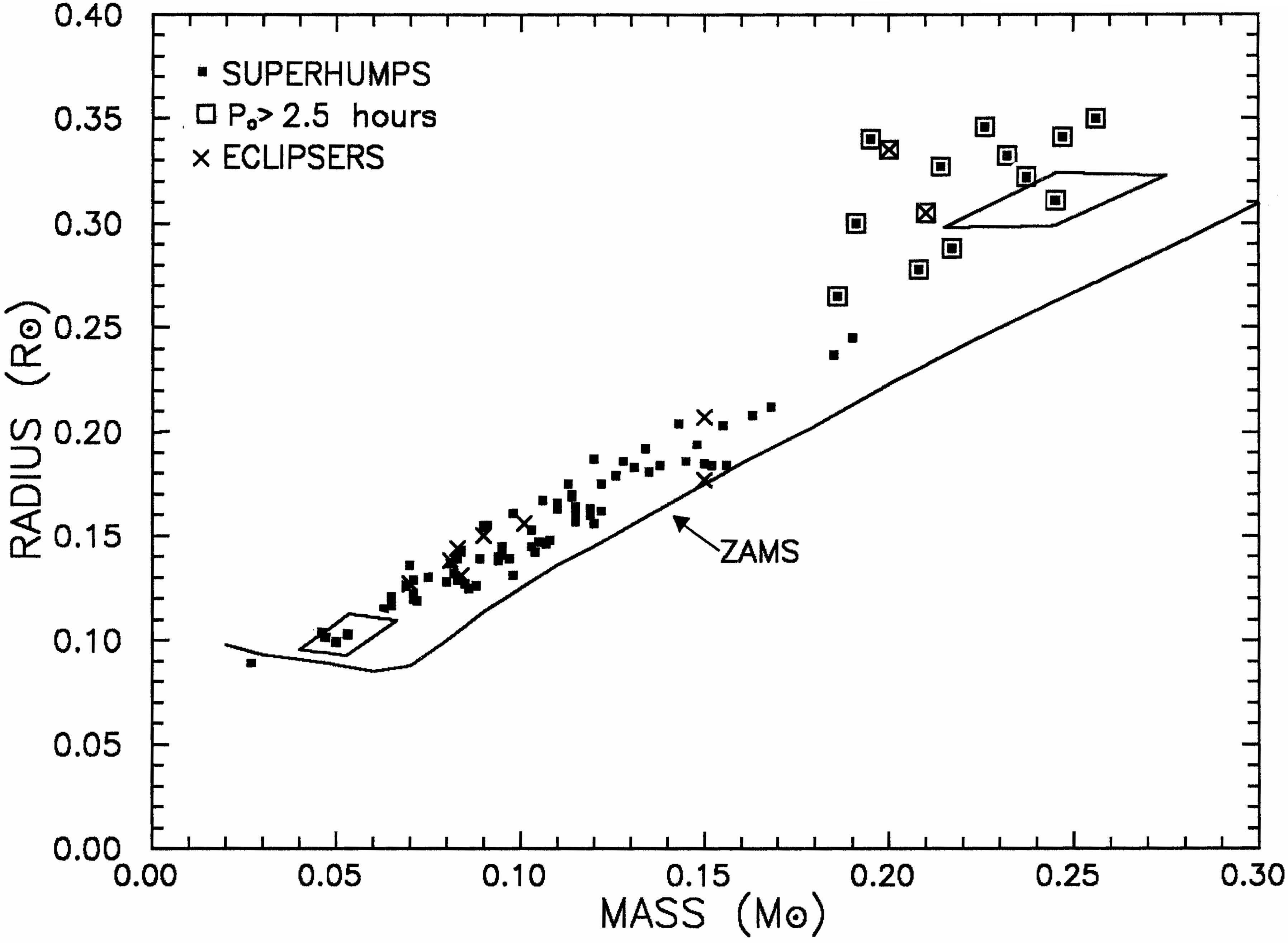

Fig 10

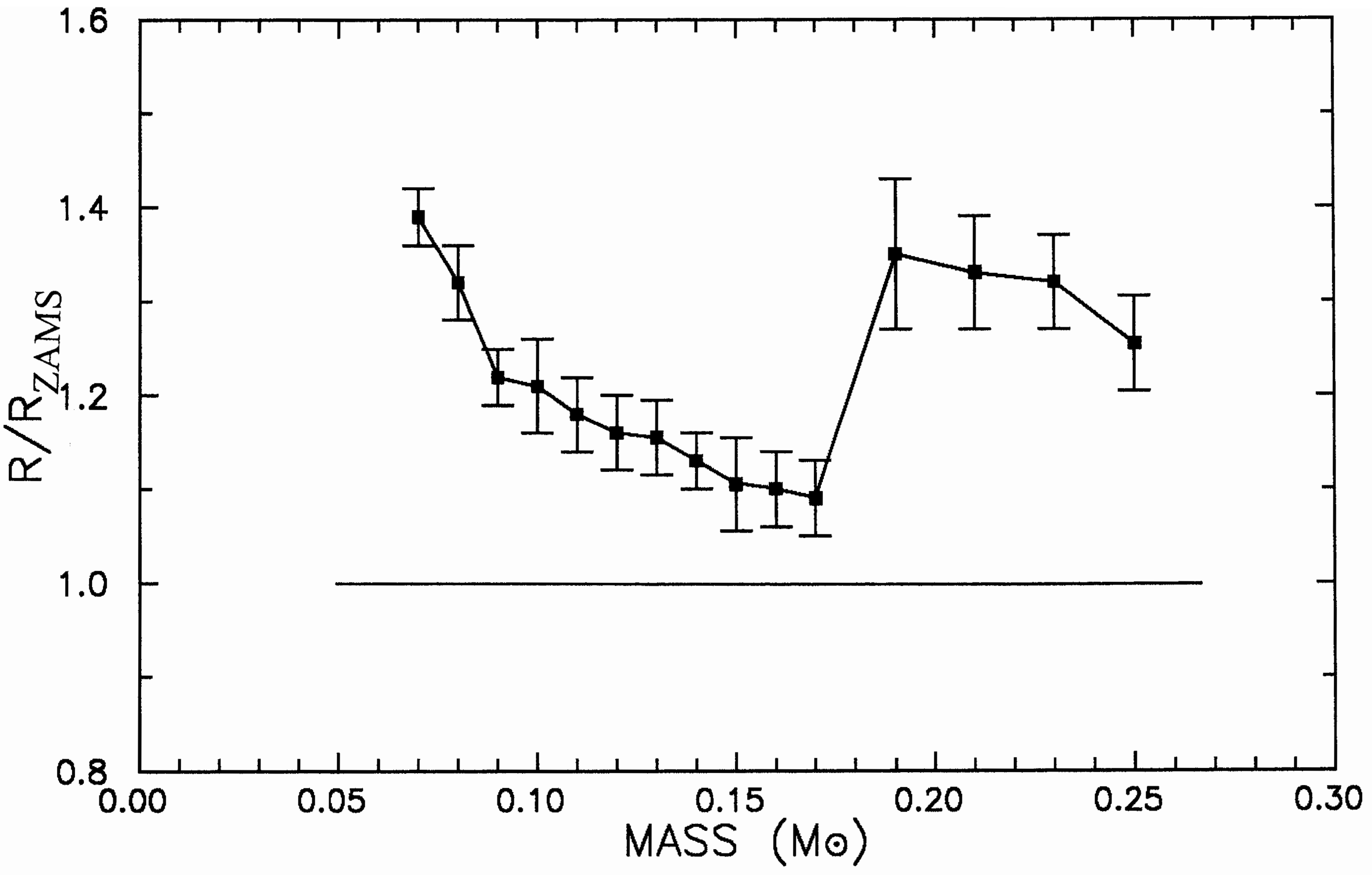

Fig 11

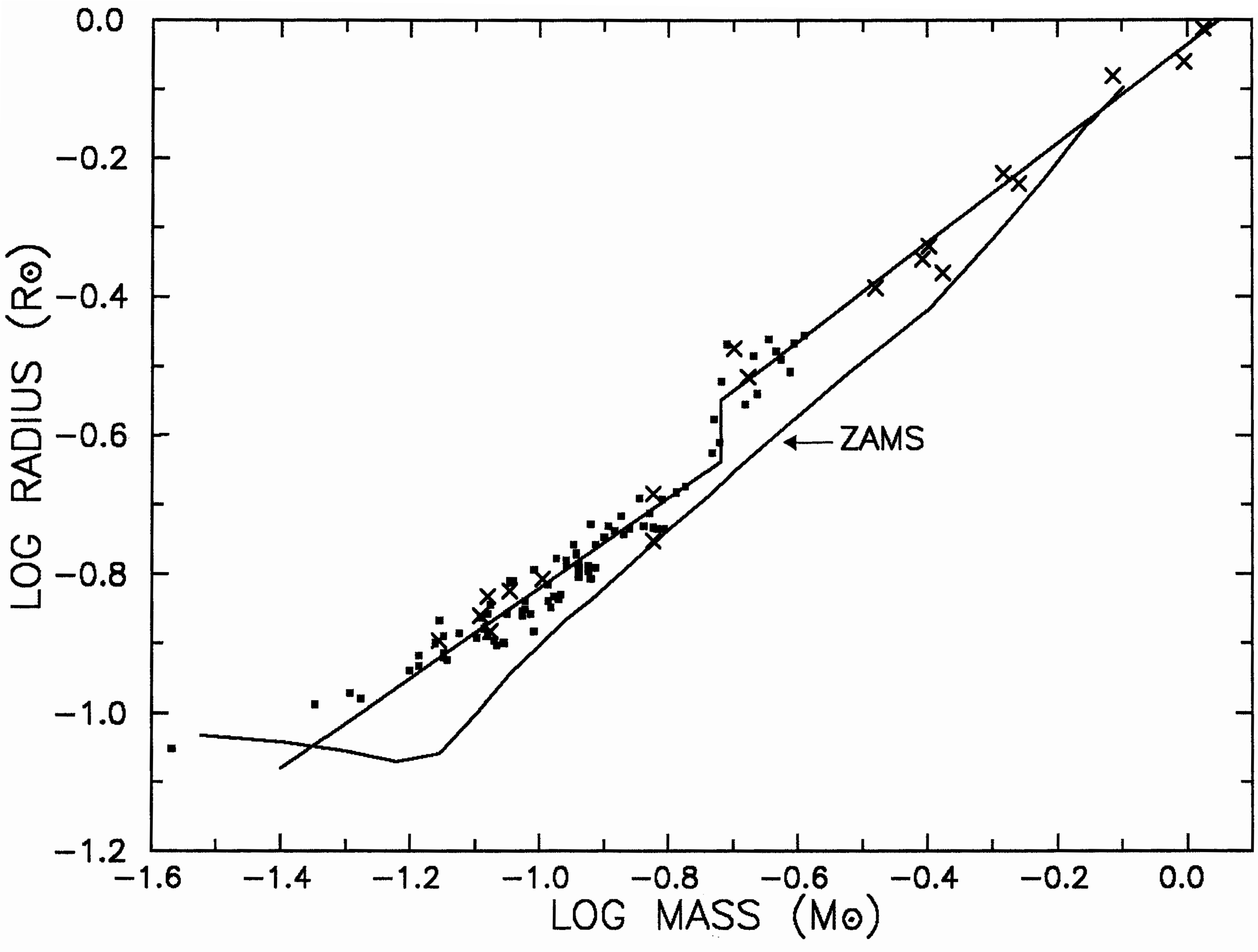

Fig 12